\documentclass{article}

\usepackage{PRIMEarxiv}

\usepackage[utf8]{inputenc} 
\usepackage[T1]{fontenc}    
\usepackage{hyperref}       
\usepackage{url}            
\usepackage{booktabs}       
\usepackage{amsfonts}       
\usepackage{nicefrac}       
\usepackage{microtype}      
\usepackage{lipsum}
\usepackage{fancyhdr}       
\usepackage{graphicx}       
\graphicspath{{media/}}     
\usepackage{float}
\usepackage{bm}
\usepackage{amsmath}

\title{Physics-guided deep reinforcement learning for flow field denoising}

\author{
  Mustafa Z. Yousif, Meng Zhang, Yifan Yang, Haifeng Zhou, Linqi Yu and HeeChang Lim\\
 School of Mechanical Engineering, Pusan National University\\
 2, Busandaehak-ro 63beon-gil,Geumjeong-gu, Busan 46241, Republic of Korea\\ 
\texttt{hclim@pusan.ac.kr} \\
}

\begin{document}
\maketitle

\begin{abstract}
A multi-agent deep reinforcement learning (DRL)-based model is presented in this study to reconstruct flow fields from noisy data. A combination of reinforcement learning with pixel-wise rewards, physical constraints represented by the momentum equation and the pressure Poisson equation and the known boundary conditions is utilised to build a physics-constrained deep reinforcement learning (PCDRL) model that can be trained without the target training data. In thePCDRL model, each agent corresponds to a point in the flow field and it learns an optimal strategy for choosing pre-defined actions. The proposed model is efficient considering the visualisation of the action map and the interpretation of the model operation. The performance of the model is tested by utilising direct numerical simulation-based synthetic noisy data and experimental data obtained by particle image velocimetry. Qualitative and quantitative results show that the model can reconstruct the flow fields and reproduce the statistics and the spectral content with commendable accuracy. Furthermore, the dominant coherent structures of the flow fields can be recovered by the flow fields obtained from the model when they are analyzed using Proper orthogonal decomposition and dynamic mode decomposition. This study demonstrates that the combination of DRL-based models and the known physics of the flow fields can potentially help solve complex flow reconstruction problems, which can result in a remarkable reduction in the experimental and computational costs.
\end{abstract}


\maketitle
\section{Introduction} \label{sec:Introduction}
The understanding of fluid flows plays a crucial role in life (for instance, in medicine, construction, transportation, aerospace and astronomy). However, fluid flow problems are usually complex with high non-linear behaviour, especially turbulent flows, which occur at generally high Reynolds numbers. In most cases, data from experiments and simulations are utilised to understand and describe the behaviour of fluids with various accuracy levels that are related to the experimental and numerical setups. Numerous methods have been developed to improve the accuracy and practicality of the obtained flow fields. However, several limitations still exist. One of the most notable limitations of the experimental approach is the noise of the obtained flow fields due to the experimental setup. Herein, obtaining measurements with an acceptable signal-to-noise ratio (SNR) is practically impossible in some cases. Therefore, several methods for the reconstruction of flow fields have been introduced. Methods based on Linear data-driven approaches, such as proper orthogonal decomposition \cite{Lumley1967} and dynamic mode decomposition \cite{Schmid2010}, have shown their capability to enhance the resolution of the flow data and filter noisy flow data \cite{Gunes&Rist2007, He&Liu2017,Fathietal2018,Nonomuraetal2019, Scherletal2020}. Additionally, various denoising methods for particle image velocimetry (PIV) measurements, such as convolution filters, wavelet methods and the Wiener filters, have revealed various levels of success \cite{Veteletal2011}. All the aforementioned methods showed limited success in terms of denoising of flow fields because they are based on linear mapping or handcrafted filtering processes, which are mostly incapable of dealing with highly non-linear fluid problems \cite{Bruntonetal2020}.

With the recent rapid development in machine learning (ML) and graphic processing unit, new data-driven methods have been introduced to provide efficient solutions for problems in various fields, such as image processing, natural language processing, robotics and weather forecasting. Several ML algorithms have been recently utilised to address problems in fluid dynamics and have shown promising results \cite{Bruntonetal2020, Duraisamyetal2019, Vinuesa&Brunton2022}. In contrast to linear methods, ML-based techniques can deal with complex non-linear problems. This feature has paved the way to explore the feasibility of applying ML to various problems in complex turbulent flows \cite{Guastonietal2021,  Yousifetal2023b}. Several supervised and unsupervised ML-based methods have been introduced considering flow reconstruction from spatially limited or corrupted data \cite{Discetti&Liu2022}. Recently promising results have been reported using deep learning (DL) by applying end-to-end trained convolutional neural network (CNN)-based models \cite{Fukamietal2019, Liuetal2020} and generative adversarial network (GAN)-based models \cite{ Yousifetal2023a, Kimetal2021, Yuetal2022}, where deep learning is a subset of machine learning, in which neural networks with multiple layers are utilised in the model \cite{LeCunetal2015}. Herein, the GAN-based models have shown better performance than the traditional CNN-based models. Nonetheless, the drawback of such methods lies in the need for the target (high-resolution or uncorrupted) flow data to train the model, which is difficult or impossible to obtain in most cases. Therefore, attempts have been recently made to address this issue under certain conditions; for instance, in the case of super-resolution reconstruction of randomly seeded flow fields \cite{Guemesetal2022} or applying physical constraints in the loss function of the model to reconstruct high-resolution steady flows from low-resolution noisy data \cite{Gaoetal2021}. However, the insufficient explainability and interpretability are the main concerns of using ML-based methods, where no concrete explanation nor control of the model performance is available.

On the other hand, Reinforcement Learning (RL), which is an ML method where an agent learns to make decisions by interacting with an environment has shown remarkable results in areas like robotics, game playing, and optimization problems \cite{Thomasetal2022}. In RL The agent takes actions and receives feedback in the form of rewards or penalties. Over time, it aims to learn the optimal actions to maximize cumulative rewards and achieve its objectives through trial and error. This approach to learning makes deep reinforcement learning (DRL) a good candidate to be applied to several problems in fluid dynamics, such as flow control \cite{Rabaultetal2019}, design optimisation \cite{Viqueratetal2021}, computational fluid dynamics \cite{Novatietal2021} and others \cite{Garnieretal2021, Viqueratetal2022}.

This paper presents a DRL-based approach that can be utilised for reconstructing flow fields from noisy data. The main advantages of the presented model lie in overcoming the necessity of the target data in the training process and the explainable filtering process of the noisy data.

The remainder of this paper is organised as follows. Section~\ref{sec:Methodology} explains the reconstruction methodology of denoised flow fields using the proposed DRL model. Section~\ref{data} describes the generation and preprocessing of the data used for training and testing the model. Section~\ref{Results and discussion} discusses the results of testing the proposed model. Finally, the conclusions of this study are presented in Section~\ref{Conclusions}.

\section{Methodology}\label{sec:Methodology}
Different from supervised and unsupervised learning, reinforcement learning is based on the Markov decision process, which is an iterative process where an agent interacts with an environment. This process comprises four elements: state $s$, action $a$, policy $\pi(a|s)$ and reward $r$. The action is an operation that is applied by the agent. The policy represents the action selection strategy of the agent. In other words, at each iteration step, the agent obtains a state and chooses an action according to the policy. Owing to the taken action, the state in the environment is then changed and the agent receives an immediate reward, which is feedback showing the usefulness of the taken action. The agent learns experience from the collected states, actions and rewards after several iterations to find an optimal policy $\pi^*(a|s)$ that maximises the long-term reward. In DRL, a deep neural network is utilised to obtain the optimal policy.

\begin{figure}
  \centerline{\includegraphics[scale=0.2]{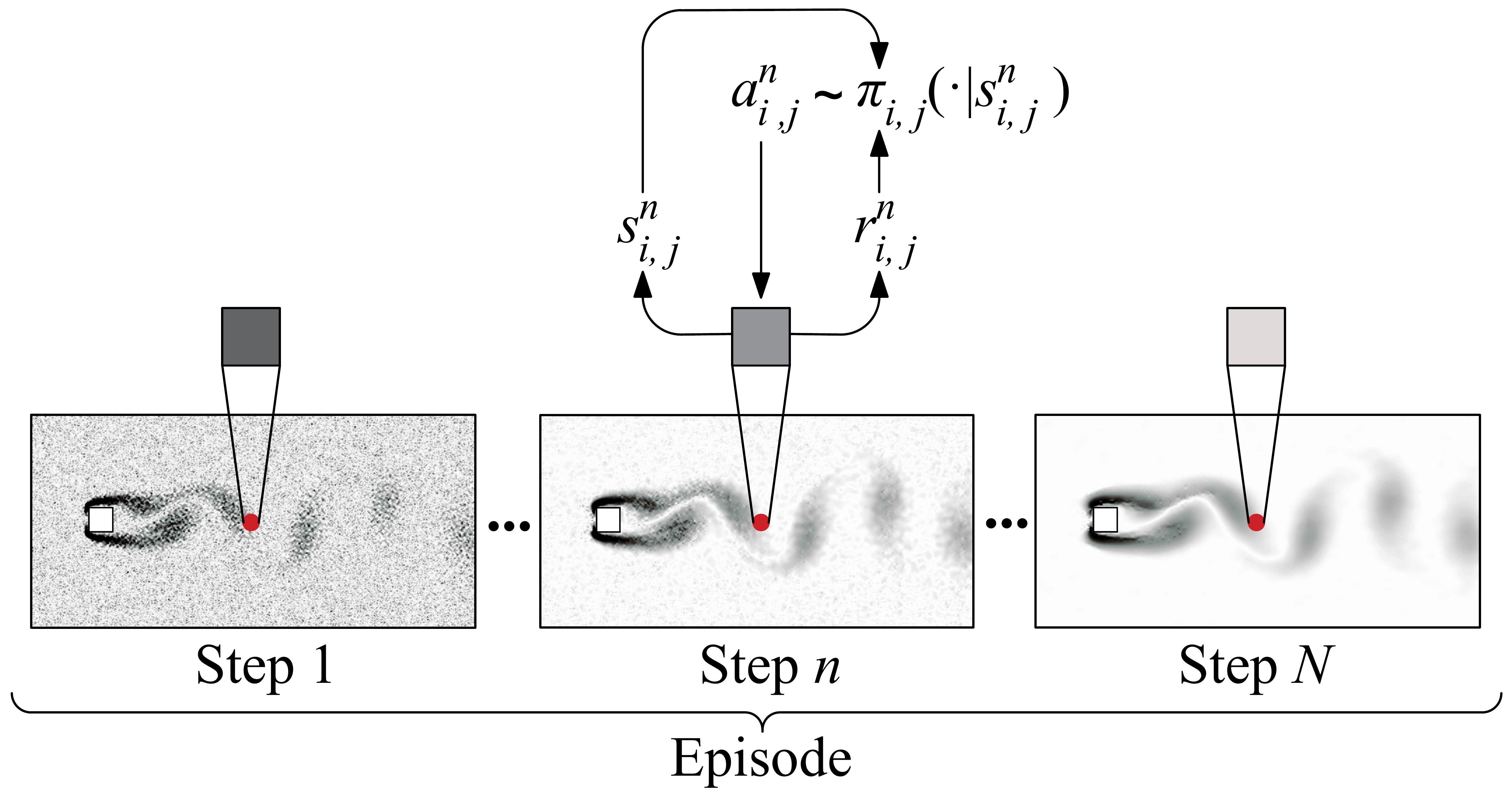}}
  \caption{Learning process in the PCDRL model. Each agent at each iteration step in the episode obtains a state from a point in the flow, calculates the reward and applies an action according to the policy.}
\label{fig:F1}
\end{figure}

This study presents a physics-constrained deep reinforcement learning (PCDRL) model that is built on the reinforcement learning with pixel-wise rewards (PixelRL) \cite{Furutaetal2020}, which is a CNN-based multi-agent DRL method for image processing \cite{Jarosiketal2021, Lietal2020, Vassiloetal2020}. In PixelRL, the asynchronous advantage actor–critic (A3C) algorithm \cite{Mnihetal2016} is applied for learning policies, which determine the actions that are represented by the choice of basic filters for each pixel. In other words, each pixel has one agent in PixelRL. A model that applies optimal policies to change the velocity values is investigated in this study by choosing the suitable actions for each point in the flow field in an instant, as shown in Figure~\ref{fig:F1}. In contrast to image processing problems that require the target data in the training process \cite{Furutaetal2020}, the physics of the flow represented by the governing equations and the known boundary conditions are utilised to train the model. 

Let $\chi^{n}_{i,j}$ be the value of an instantaneous velocity component at iteration step $n$ and in the location $(i,j)$ of the field. Herein, each location has its agent with a policy $\pi_{i,j}(a^{n}_{i,j}|s^{n}_{i,j})$, where $a^{n}_{i,j}\in \mathcal A$, which is a pre-defined set of actions (Appendix~\ref{appD}). Each agent obtains the next state, that is, $s^{n+1}_{i,j}$ and reward $r^{n+1}_{i,j}$ from the environment by taking the action $a^{n}_{i,j}$.

Physical constraints represented by the momentum equation and the pressure Poisson equation and the known boundary conditions are embedded in the reward function, which enables the model to follow an optimal denoising strategy that results in changing the noisy data to the true flow field distribution. Hence, the objective of the model is to learn the policy that maximises the expected long-term rewards:
\begin{equation} \label{eqn:eq1}
\pi^{*}_{i,j} = \mathop{\mathrm{argmax}}\limits_{\pi_{i,j}} E_{\pi_{i,j}} (\displaystyle\sum_{n=1}^{N}\gamma^{(n-1)}r^{n}_{i,j}),
\end{equation}
\noindent where $\gamma^{(n-1)}$ is the $(n-1)$-th power of the discount factor $\gamma$, which determines the weights of the immediate rewards in the iteration steps. In this study, the value of $\gamma$ is set to 0.95.

The combination of the momentum equation, 
\begin{equation} \label{eqn:eq2}
\frac{\partial \textbf{\textit{u}}}{\partial t} + (\textbf{\textit{u}}\cdot \bm{\nabla})\textbf{\textit{u}} = - \bm{\nabla} p + \nu \bm{\nabla}^2\textbf{\textit{u}},
\end{equation}
\noindent and the pressure Poisson equation,
\begin{equation} \label{eqn:eq3}
 \bm{\nabla} \cdot (\textbf{\textit{u}}\cdot \bm{\nabla})\textbf{\textit{u}} = - \bm{\nabla}^2 p ,
\end{equation}
\noindent is utilised to build the physics-based immediate reward, $(r^{n}_{i,j})_{Physics}$, where $\textbf{\textit{u}}$, $p$, $t$ and $\nu$ are the velocity vector, pressure (divided by density), time and kinematic viscosity, respectively.

At each iteration step, the pressure field is obtained by numerically solving equation~(\ref{eqn:eq3}). Herein, the pressure gradient calculated from the pressure field  (($ \bm{\nabla}{p}^{n}_{i,j})_{Poisson}$) is utilised in equation~(\ref{eqn:eq2}) such that

\begin{equation} \label{eqn:eq4}
(r^{n}_{i,j})_{Physics} = -|(\frac{\partial \textbf{\textit{u}}}{\partial t} + (\textbf{\textit{u}}\cdot\bm{\nabla})\textbf{\textit{u}}- \nu \bm{\nabla}^2\textbf{\textit{u}})^{n}_{i,j}+( \bm{\nabla} {p}^{n}_{i,j})_{Poisson}|.
\end{equation}

The pressure integration in equation~(\ref{eqn:eq3}) is done by utlising a Poisson solver that applies a standard 5-point scheme (second-order central difference method)\cite{VanderKindereetal2019} with the initial pressure field being estimated from numerically integrating the pressure gradient obtained from the initial noisy data in equation~(\ref{eqn:eq2}) \cite{Oudheusdenetal2007}. Notably, the central difference method is applied for all the spatial discretisations. Regarding the temporal discretisation, for the first and the last time steps in each training mini-batch, the forward difference and the backward difference are used, respectively, and the central difference is applied for the other time steps. Furthermore, Neumann and Dirichlet boundary conditions according to each case used in this study are enforced in the calculations. 

Additionally, the velocity values obtained after each action $a^{n}_{i,j}$ are directly made divergence-free by applying Helmholtz–Hodge decomposition \cite{Bhatiaetal2013} using Fourier transformation. Furthermore, the known boundary conditions are utilised to obtain the boundary conditions-based immediate reward $(r^{n}_{i,j})_{BC}$ for the velocity by considering the absolute error of the reconstructed data at the boundaries of the domain. 

Thus, the combined immediate reward function can be expressed as

\begin{equation} \label{eqn:eq5}
r^{n}_{i,j} = (r^{n}_{i,j})_{Physics} + \beta(r^{n}_{i,j})_{BC},
\end{equation}
\noindent where $\beta$ is a weight coefficient and its value is empirically set to 20.

This approach considers the convergence of the model output to satisfy the governing equations and boundary conditions as a measure of the model performance without the need for the target training data. Furthermore, the reward function is designed to mimic the denoising process of PIV velocity field data without the need for measured pressure field data in the model. Nine iteration steps for each episode, that is, $N = 9$, are used in this study. In addition, the size of the training mini-batch is set to 4. The model is applied to direct numerical simulation (DNS)-based data (corrupted by different levels of additive zero-mean Gaussian noise) and real noisy PIV data of two-dimensional flow around a square cylinder at Reynolds number, $Re_D$ = 100 and 200, respectively. Herein, $Re_D=u_{\infty} D/\nu$, where $u_{\infty}$ and $D$ are the free-stream velocity and the cylinder width, respectively. Details regarding the source code of the proposed model, A3C, PixleRL and the selected pre-defined denoising action set can be found in appendices \ref{appA}, \ref{appB}, \ref{appC} and \ref{appD}, respectively.

\section{Data description and preprocessing}\label{data}
\subsection{Synthetic data}

DNS data of a two-dimensional flow around a square cylinder at a Reynolds number, $Re_D=100$, are considered as an example of synthetic data. The open-source computational fluid dynamics finite-volume code OpenFOAM-5.0x is used to perform the DNS. The domain size is set to be $(x_D\times y_D) = (20\times15)$, where $x$ and $y$ are the streamwise and spanwise directions, respectively. The corresponding grid size = $(381\times221)$. Local mesh refinement is applied using the stretching mesh technique near the cylinder walls. Uniform inlet velocity and pressure outlet boundary conditions are applied to the inlet and the outlet of the domain, respectively. No-slip boundary condition is applied to the cylinder walls and the symmetry plane to the sides of the domain. The dimensionless time step of the simulation, that is, $u_{\infty} \Delta t/D$, is set to $10^{-2}$. The DNS data are corrupted by additive zero-mean Gaussian noise, that is, $\mathcal S\sim \mathcal N (0,\sigma^2)$, where $\mathcal S$, $\mathcal N$ and $\sigma^2$ represent the noise, the normal distribution and the variance, respectively. The signal-to-noise ratio (SNR), wherein a large SNR yields a low noise level, is used to evaluate the noise level. Herein, SNR$ =\sigma^2_{DNS}/\sigma^2_{noise}$, where $\sigma^2_{DNS}$ and $\sigma^2_{noise}$ denote the variance of the DNS and the noise data, respectively. Three levels of noise are applied, $1/{\rm SNR} =$ 0.01, 0.1 and 1. The interval between the collected snapshots of the flow fields is set to be 10 times the simulation time step; 1000 snapshots are used for training the model, whereas 200 snapshots are used for testing the performance of the model.

\subsection{Experimental data}
Two PIV experiments are performed to generate noisy and clear (uncorrupted) data (for comparison) of flow over a square cylinder to investigate the performance of the proposed PCDRL model on real experimental data. The noisy data are generated by utilising a return-type water channel. The test section size of the water channel is 1 m (length) $\times$ 0.35 m (height) $\times$ 0.3 m (width). The free-stream velocity is set to 0.02 m/s, with the corresponding $Re_D$ of 200. The background noise is generated at relatively high levels due to the external noise and the sparse honeycomb of the water channel. The channel was seeded by polyamide12 seed particles from INTECH SYSTEMS with 50 $\mu$m diameter. The high-speed camera (FASTCAM Mini UX 50) and a continuous laser with a 532 nm wavelength are utilised to build the complete PIV system. The snapshot frequency is set to 24 Hz. Herein, 2,000 and 500 instantaneous flow fields are used for the model training and testing of its performance, respectively. Meanwhile, clear data of the flow are generated by utilising a return-type wind tunnel. The test section size of the wind tunnel is 1 m (length) $\times$ 0.25 m (height) $\times$ 0.25 m (width). The free-stream velocity is set to 0.29 m/s, with the corresponding $Re_D$ of 200. The turbulence intensity of the free-stream is less than $0.8\%$. The wind tunnel is seeded by the olive oil droplets generated by TSI 9307 particle generator. The PIV system used in the wind tunnel comprises a two-pulsed laser (Evergreen, EVG00070) and a CCD camera (VC-12MX) with $4,096 \times 3,072$ pixel resolution. Herein, the snapshot frequency is set to 15 Hz. In the water channel and wind tunnel experiments, the square cylinder model comprised an acrylic board and the cross-section of the model is set to 1 cm $\times$ 1 cm. The model is not entirely transparent. Thus, a shadow region is generated in the area below the bluff body when the laser goes through the model.

\section{Results and discussion} \label{Results and discussion}
\subsection{Performance of the model}
\begin{figure}
  \centerline{\includegraphics[scale=0.25]{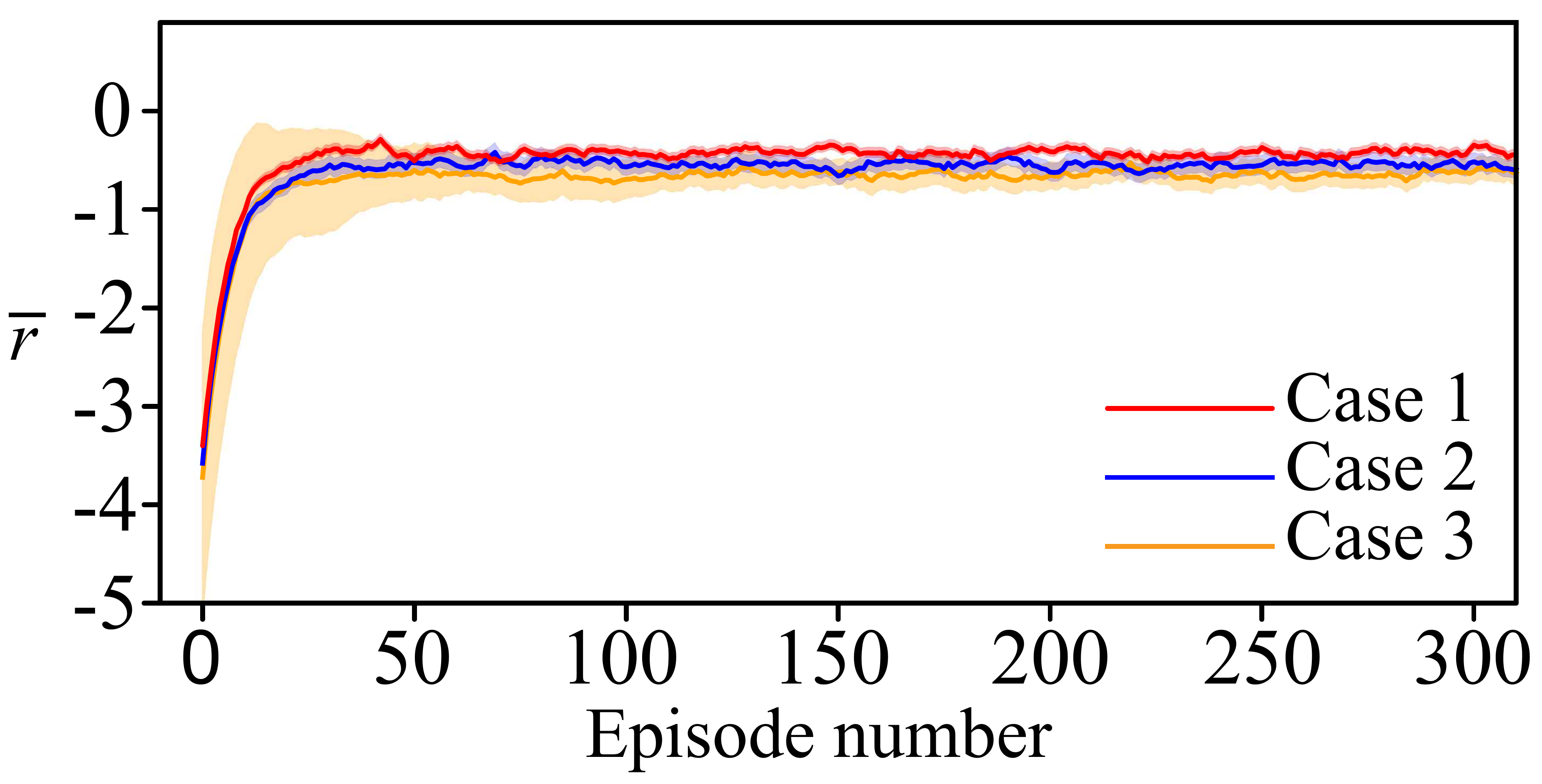}}
  \caption{Progress of the mean reward during the training process. Cases 1, 2 and 3 represent the noisy DNS data at noise levels $1/{\rm SNR} =$ 0.01, 0.1 and 1, respectively.}
\label{fig:F2}
\end{figure}

The capability of the PCDRL model to denoise flow fields is investigated in this study qualitatively and quantitatively by utilising the DNS and PIV data. The model is primarily applied to DNS-based data. Figure~\ref{fig:F2} shows the progress of the mean reward during the training process, that is,
\begin{equation} \label{eqn:eq6}
\bar{r} = \frac{1}{IJN} \sum_{i=1}^{I} \sum_{j=1}^{J} \sum_{n=1}^{N} r^{n}_{i,j}.
\end{equation}

The solid line and light area indicate $\bar{r}$ and the standard deviation of the reward on nine iteration steps, respectively. As shown in the figure, the reward for the three different noise levels rapidly increases and approaches its optimal level after a few number of episodes. This finding indicates that the agents in PixelRL learn the policy in a few episodes in the training process compared with the other multi-agent networks because they share the information represented by the network parameters and also due to the averaged gradients \cite{Furutaetal2020}. Thus, this approach can remarkedly reduce the computational cost of the model. Furthermore, as expected, the magnitude of the optimal $\bar{r}$ decreases with the increase in the noise level.

\begin{figure}
  \centerline{\includegraphics[scale=0.35]{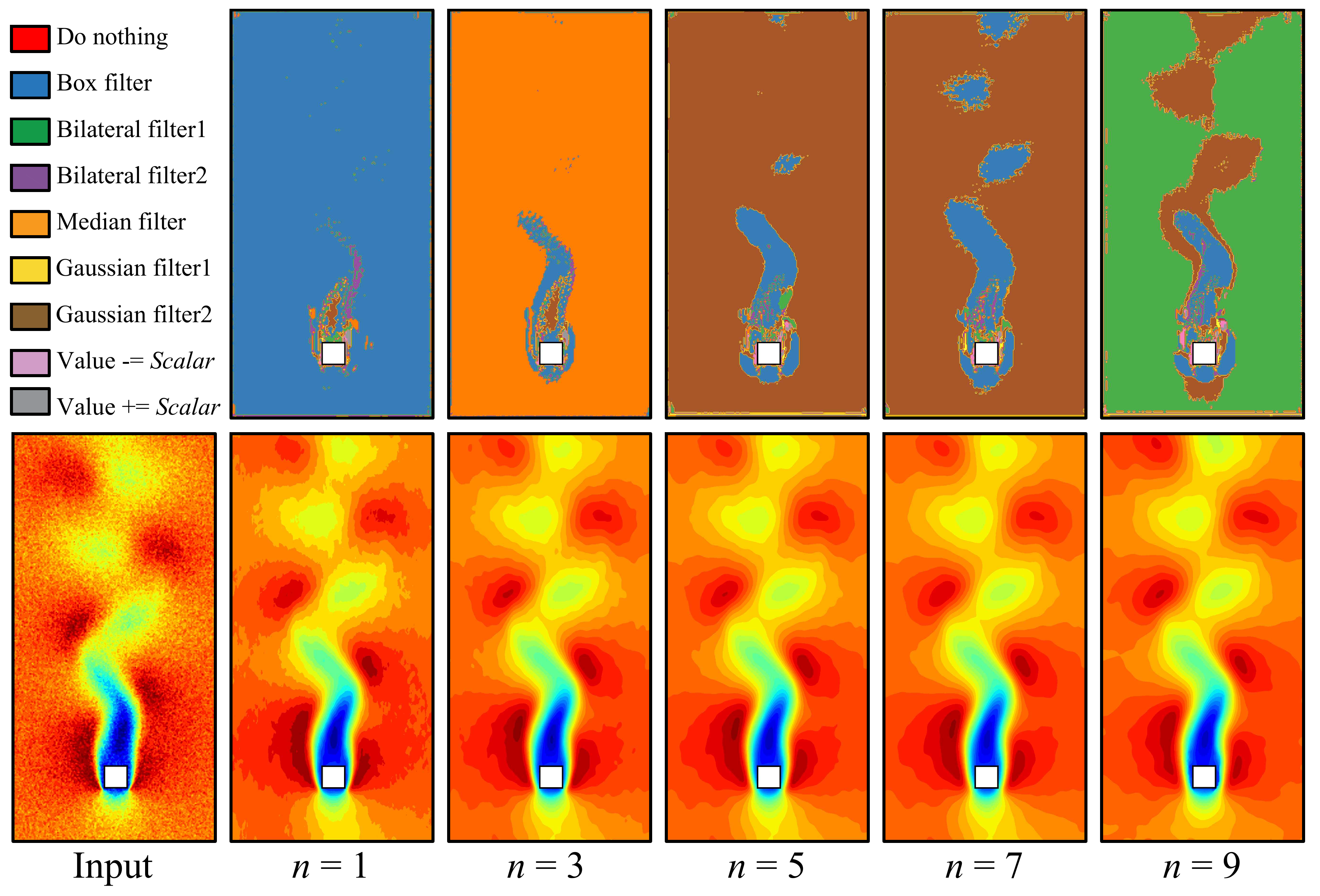}}
  \caption{Action map of the prediction process for an instantaneous streamwise velocity field. The top part of the figure shows the types of filters used in the process and the action map in each iteration step and the bottom part of the figure shows the corresponding velocity field. Result for DNS noisy data at noise level $1/{\rm SNR} =$ 0.1.}
\label{fig:F3}
\end{figure}

Figure~\ref{fig:F3} shows a visual overview of the prediction process of the PCDRL model. The figure reveals that the choice of filters changes with the spatial distribution of the velocity data and also with each iteration step in the episode. The visualisation of the action map is one of the model features, providing additional access to the model considering the action strategy. Furthermore, it can be seen that the action map is strongly correlated with the physics of the flow, which is represented in this case by the vortex shedding behind the square cylinder.

\begin{figure}
  \centerline{\includegraphics[scale=0.35]{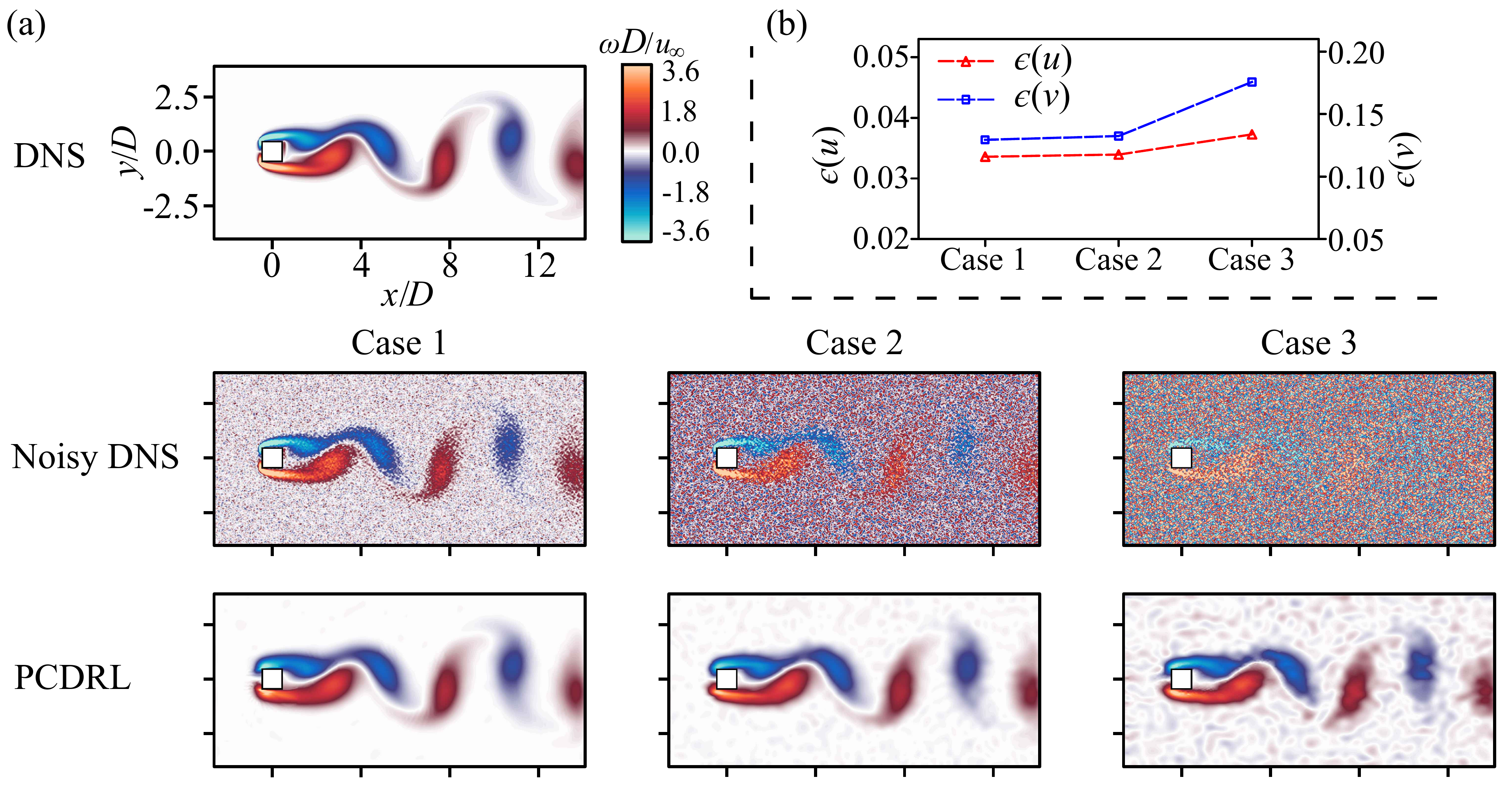}}
  \caption{(a) Instantaneous vorticity field; (b) Relative $L_2$-norm error of the reconstructed velocity fields. Cases 1, 2 and 3 represent the noisy DNS data at noise levels $1/{\rm SNR} =$ 0.01, 0.1 and 1, respectively.}
\label{fig:F4}
\end{figure}

The instantaneous denoised flow data are presented in Figure~\ref{fig:F4}(a) by employing the vorticity field ($\omega$). The figure reveals that the model shows a remarkable capability to reconstruct the flow field even when using an extreme level of noise in the input data of the model.

The general reconstruction accuracy of the model is examined via the relative $L_2$-norm error of the reconstructed velocity fields,
\begin{equation}\label{eqn:eq7}
\epsilon (\chi) = \frac{1}{K} \sum_{k=1}^{K} \frac{||\chi^{PCDRL}_k-\chi^{DNS}_k||_2}{||\chi^{DNS}_k||_2},
\end{equation}
where $\chi^{PCDRL}_k$ and $\chi^{DNS}_k$ represent the predicted velocity component and the ground truth (DNS) one, respectively, and $K$ is the number of test snapshots. Figure~\ref{fig:F4}(b) shows that the values of the error are relatively small for the velocity components and are proportional to the increase in the noise level.

\begin{figure}
  \centerline{\includegraphics[scale=0.35]{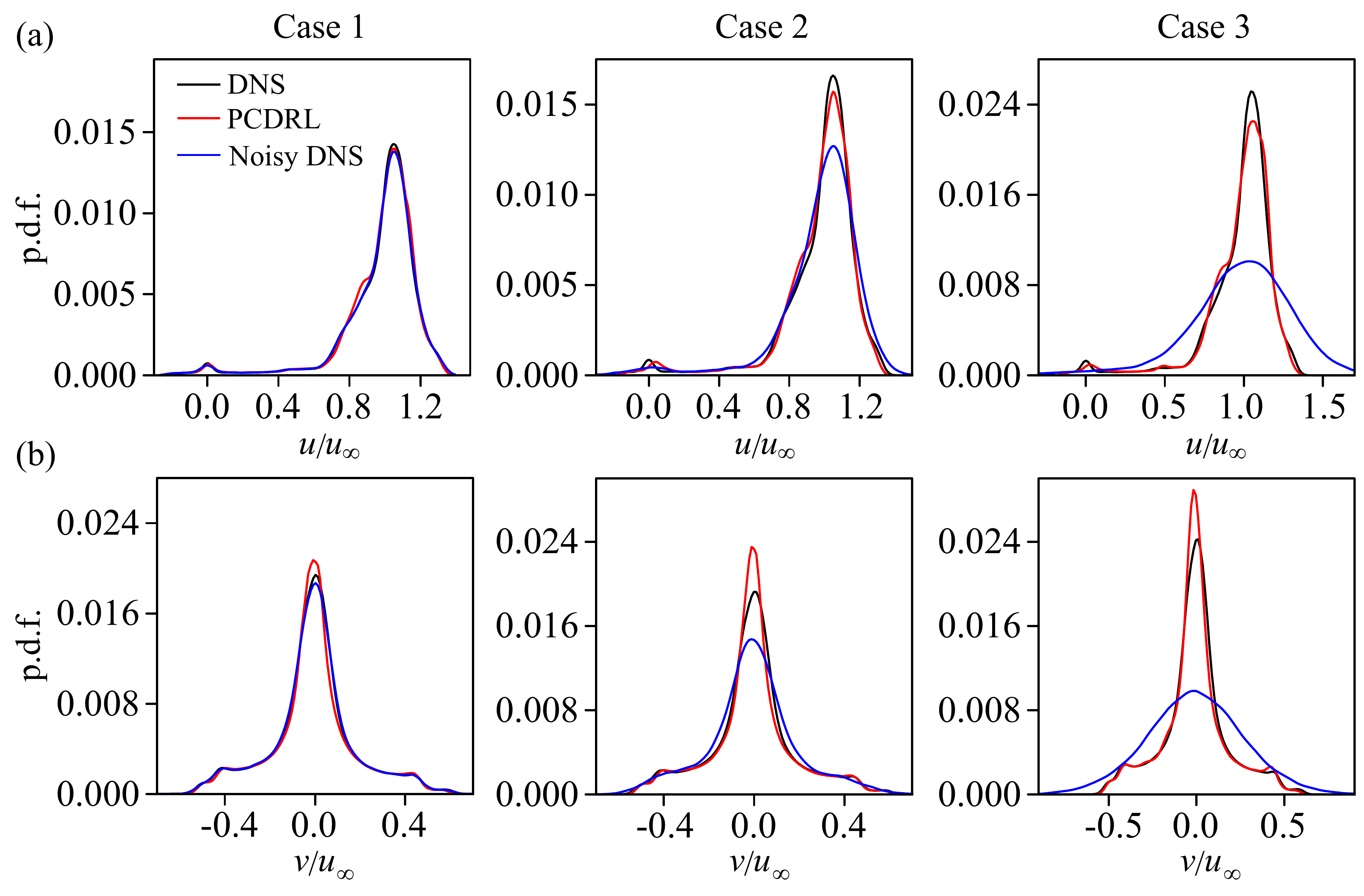}}
  \caption{Probability density function plots of the (a) streamwise and (b) spanwise velocity components. Cases 1, 2 and 3 represent the noisy DNS data at noise levels $1/{\rm SNR} =$ 0.01, 0.1 and 1, respectively.}
\label{fig:F5}
\end{figure}

\begin{figure}
  \centerline{\includegraphics[scale=0.35]{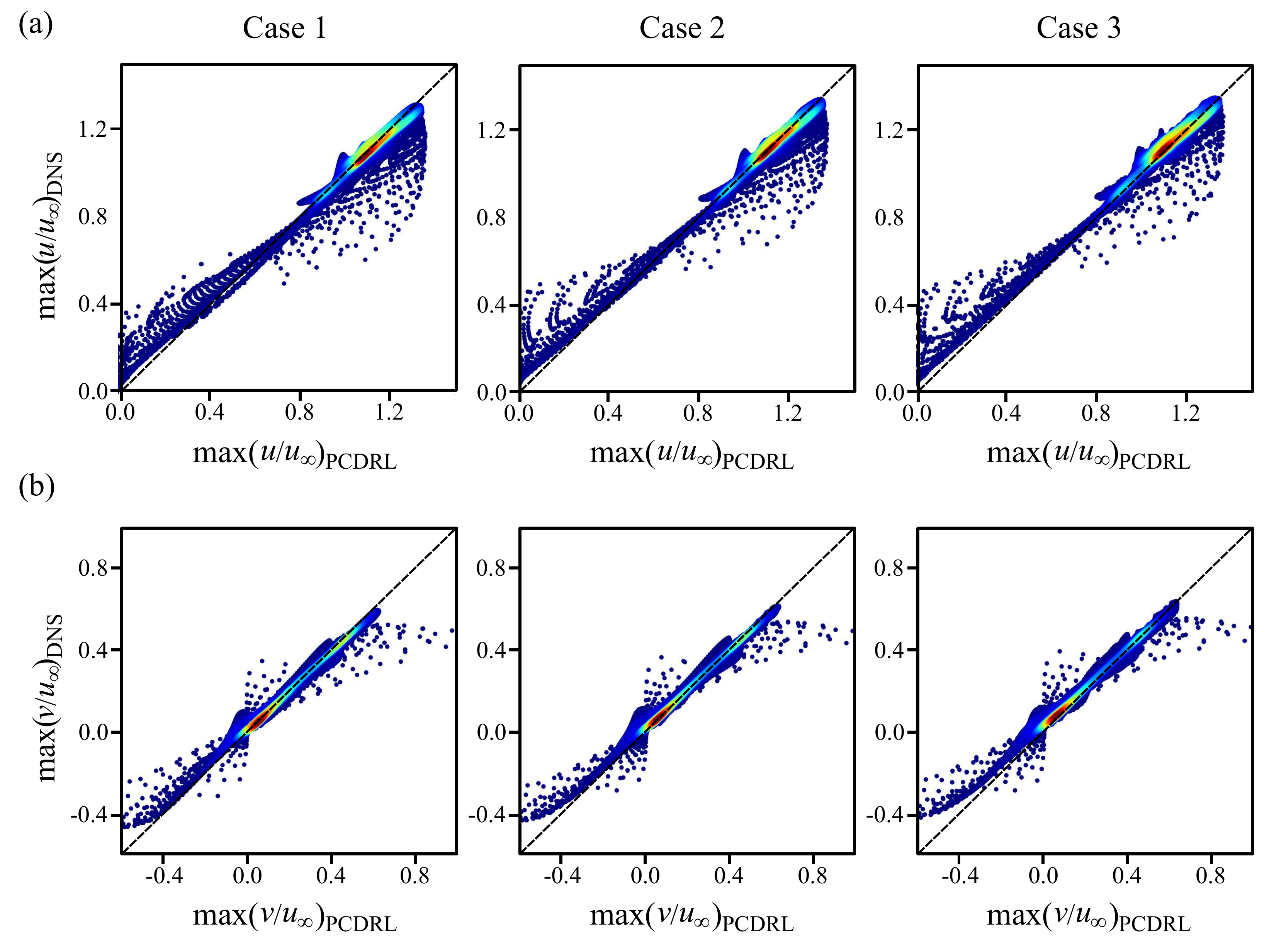}}
  \caption{Scatter plots of the maximum instantaneous values of the (a) streamwise and (b) spanwise velocity components. Cases 1, 2 and 3  represent the results from the PCDRL model using noisy DNS data at noise levels $1/{\rm SNR} =$ 0.01, 0.1 and 1, respectively. The contour colours (from blue to red) are proportional to the density of points in the scatter plot.}
  
\label{fig:F6}
\end{figure}

Figure~\ref{fig:F5} shows the probability density function (p.d.f.) plots of the streamwise ($u$) and spanwise ($v$) velocity components. Herein, the p.d.f. plots obtained from the reconstructed velocity fields are generally consistent with those obtained from DNS, indicating that the proposed model could successfully recover the actual distribution of flow data. Furthermore, the scatter plots of the maximum instantaneous velocity values in all the test data set are presented in Figure~\ref{fig:F6}. The figure reveals that the predicted data are generally in commendable agreement with the DNS data for the entire range of each velocity component with a slight reduction in the consistency as the noise level increases.

\begin{figure}
  \centerline{\includegraphics[scale=0.32]{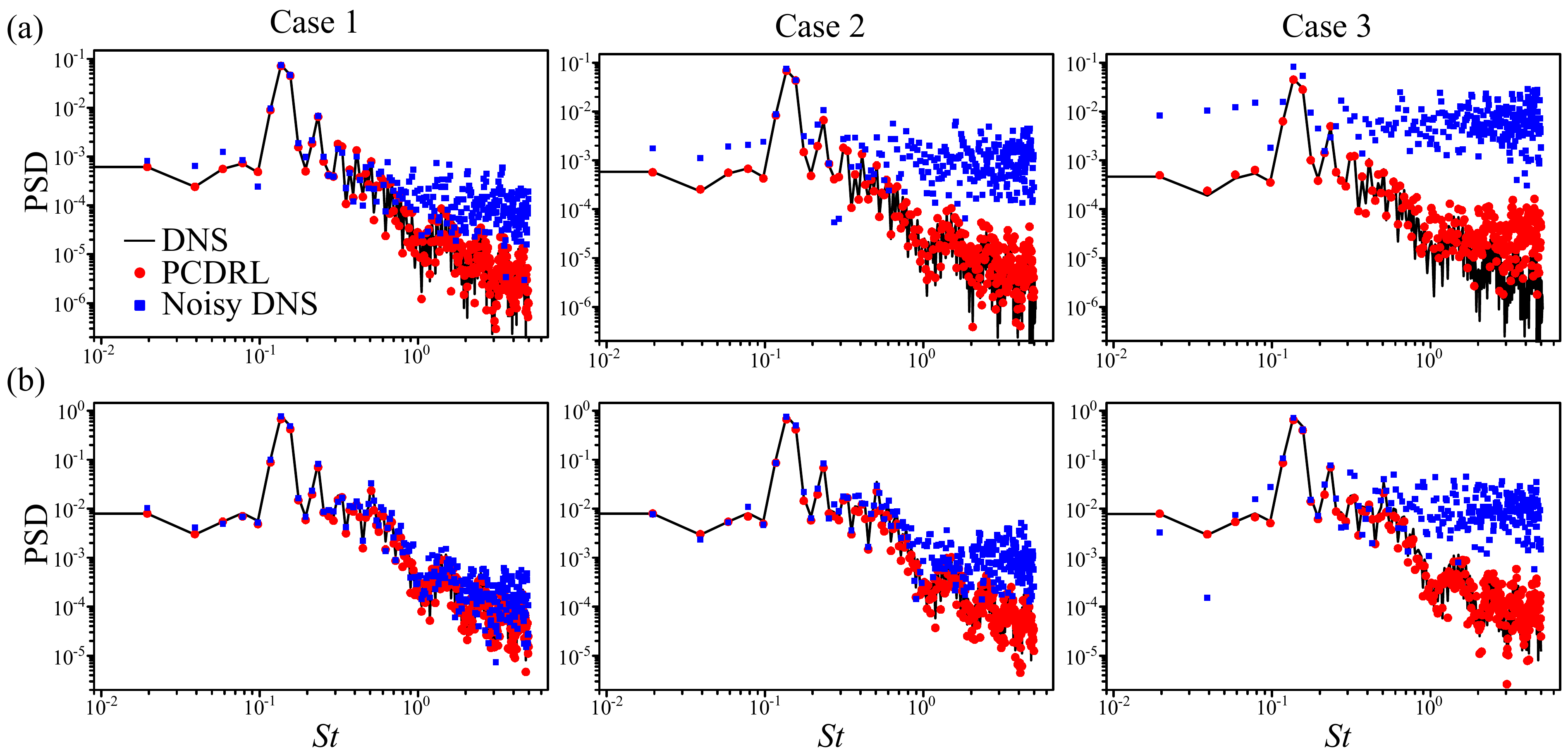}}
  \caption{Power spectral density plots of the streamwise velocity fluctuations at two different locations: (a) ($x/D,y/D$) = (1,1) and (b) ($x/D,y/D$) = (6,1). The dimensionless frequency is represented by the Strouhal number, $St=fD/u_{\infty}$, where $f$ is the frequency. Cases 1, 2 and 3 represent the noisy DNS data at noise levels $1/{\rm SNR} =$ 0.01, 0.1 and 1, respectively.}
\label{fig:F7}
\end{figure}

The power spectral density (PSD) of the streamwise velocity fluctuations at two different locations is plotted in Figure~\ref{fig:F7} to examine the capability of the model to reproduce the spectral content of the flow. Herein, a commendable agreement with the DNS results can be observed with a slight deviation in the high frequencies for the noise level, $1/{\rm SNR} = 1$.

\begin{figure}
  \centerline{\includegraphics[scale=0.5]{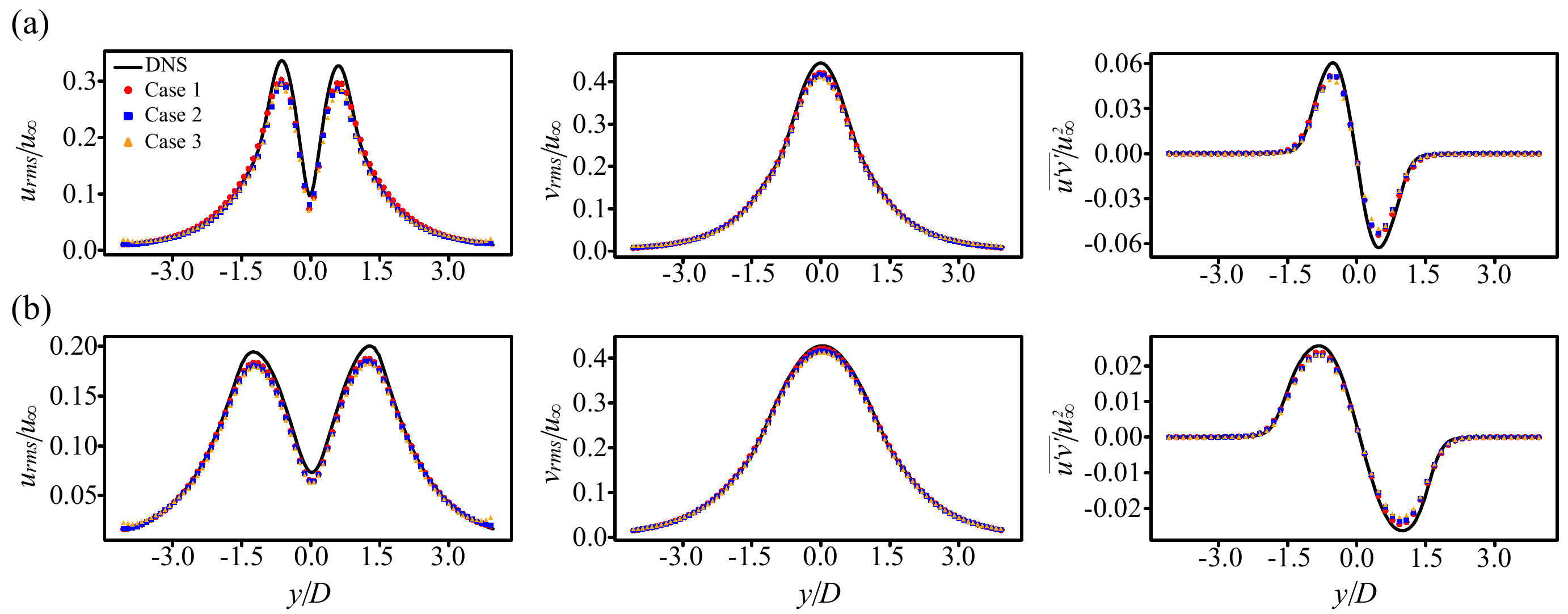}}
  \caption{Spanwise profiles of flow statistics ($u_{rms}$ (left), $v_{rms}$ (middle) and $\overline{u'v'}$ (right)) at two different streamwise locations. (a) $x/D = 3$; (b) $x/D = 6$. Cases 1, 2 and 3 represent the results from the PCDRL model using noisy DNS data at noise levels $1/{\rm SNR} =$ 0.01, 0.1 and 1, respectively.}
\label{fig:F8}
\end{figure}

The statistics of the velocity fields, represented by the spanwise profiles of the root mean square of the velocity ($u_{rms},v_{rms}$) and Reynolds shear stress ($\overline{u'v'}$), are presented in Figure~\ref{fig:F8}. The figure reveals an accurate reconstruction of the statistics at two different streamwise locations in the domain, indicating that the model could successfully reproduce the statistics of the flow despite of the extreme noise level.

\begin{figure}
  \centerline{\includegraphics[scale=0.32]{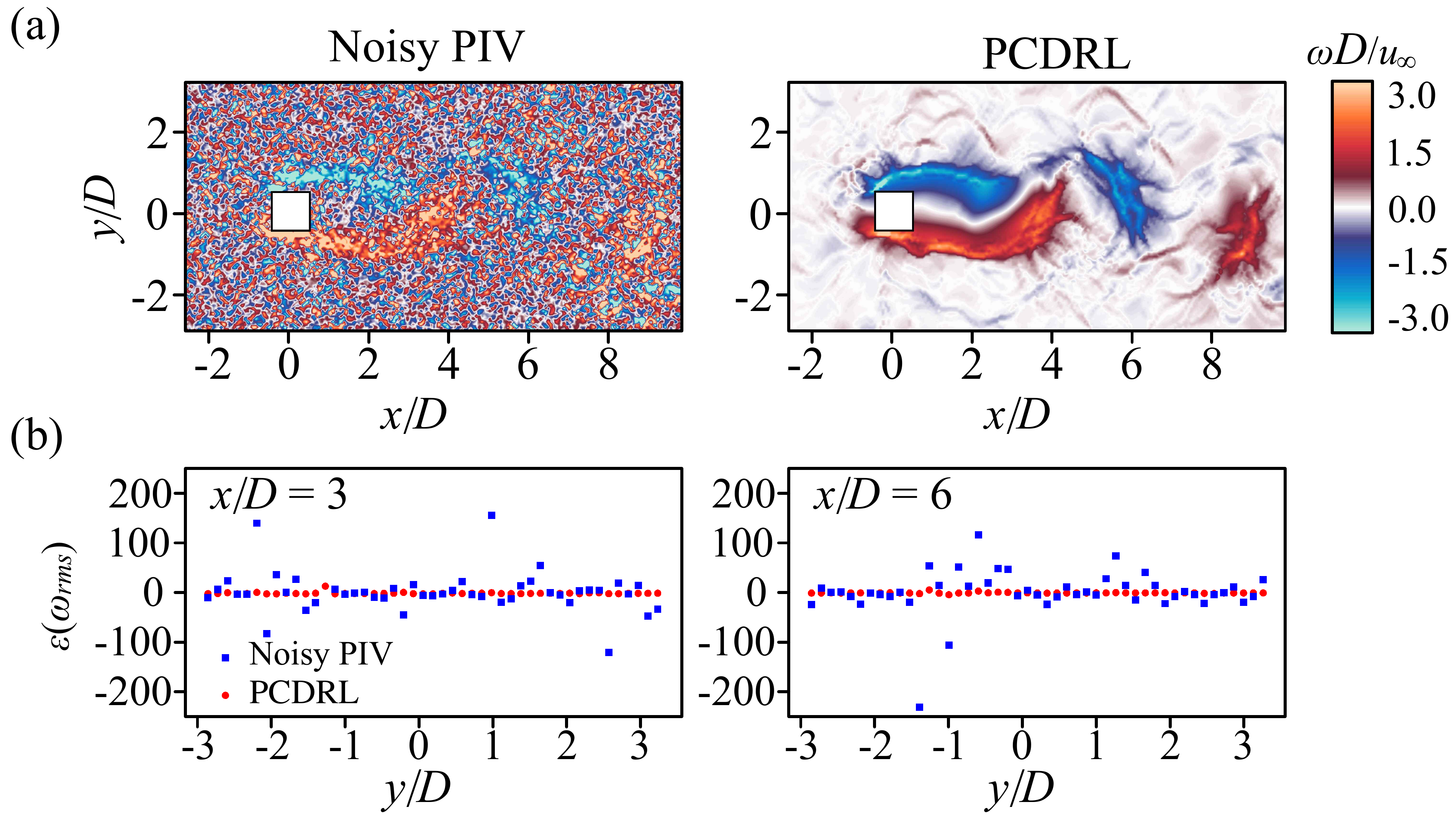}}
  \caption{(a) Instantaneous vorticity field of the noisy (left) and denoised PIV data by the PCDRL (right); (b) Relative difference of the spanwise profile of the vorticity root mean square at two different streamwise locations.}
\label{fig:F9}
\end{figure}

The model performance is further examined by utilising actual noisy PIV data. The reconstructed instantaneous vorticity field is shown in Figure~\ref{fig:F9}(a). The figure reveals that the model could successfully denoise the velocity fields with commendable accuracy considering the input noisy data to the model. In addition, the model shows the capability of recovering the corrupted regions in the flow due to the experimental setup. Furthermore, the relative difference of the spanwise profile of the vorticity root mean square ($\omega_{rms}$) between the reconstructed data and the clear PIV data ($\varepsilon(\omega_{rms})$) presented in Figure~\ref{fig:F9}(b) shows that the results from the model reveal a smooth behaviour that is generally consistent with that from the clear PIV data. These results indicate that the PCDRL model can be practically applied to noisy PIV data.

\subsection{POD and DMD results}

\begin{figure}
  \centerline{\includegraphics[scale=0.52]{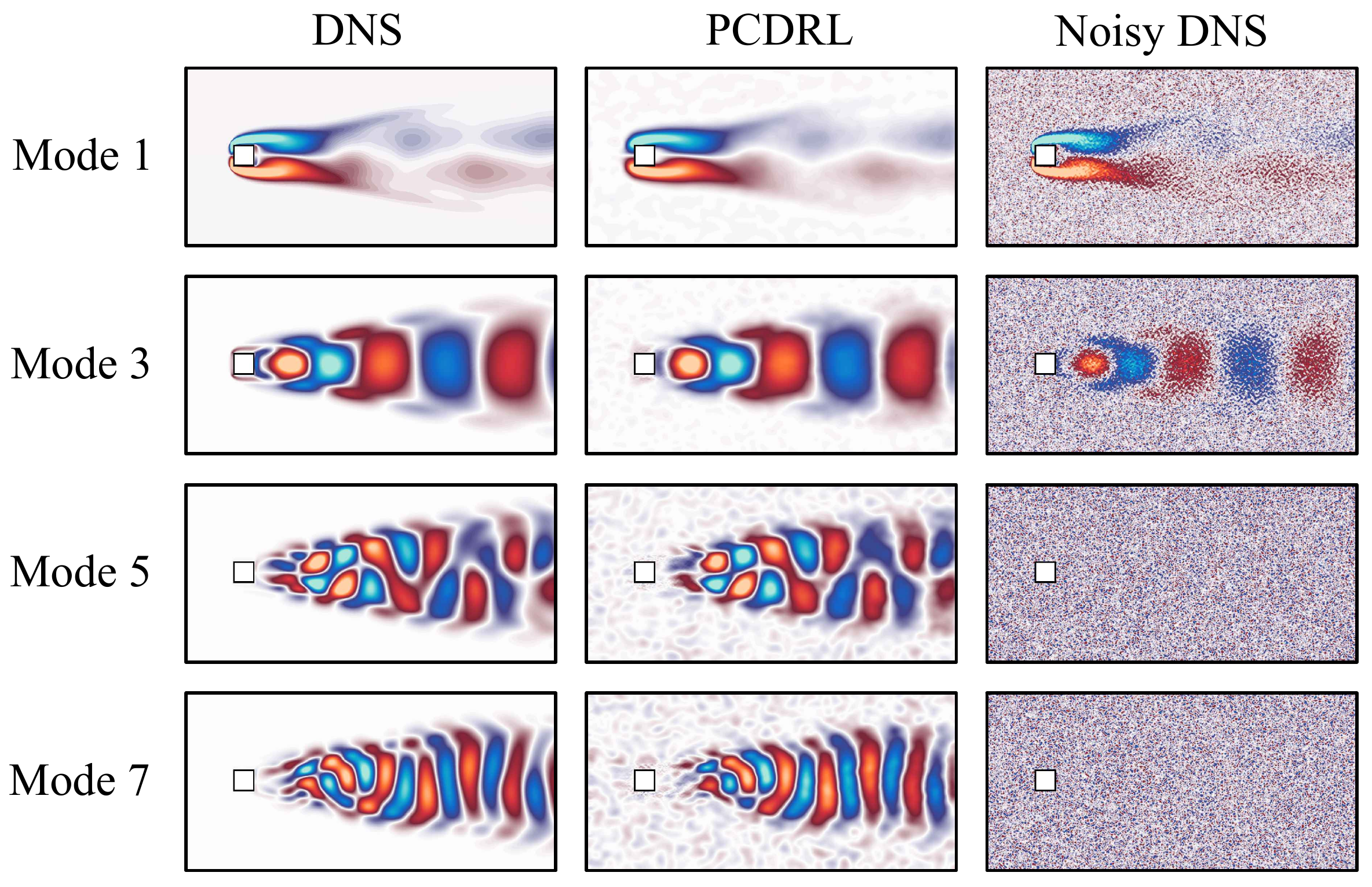}}
  \caption{Leading POD modes obtained from the DNS data. Results from ground truth DNS (left), PCDRL (middle), and noisy data with 1/SNR = 1 (right).}
\label{fig:F10}
\end{figure}

\begin{figure}
  \centerline{\includegraphics[scale=0.4]{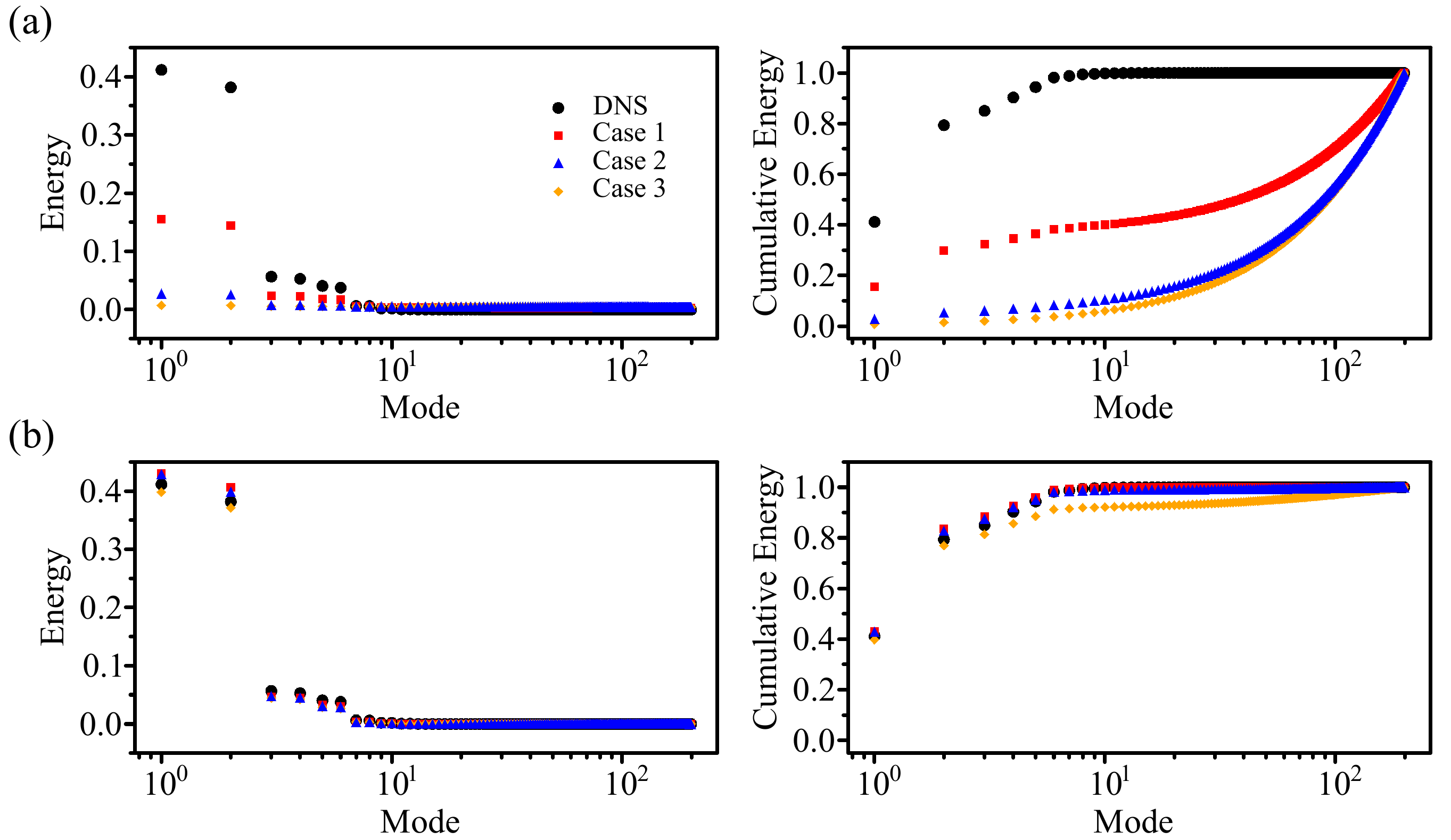}}
  \caption{Normlised energy (left) and cumulative energy (right) of the POD modes obtained from the DNS data; (a) Noisy data, where Cases 1, 2 and 3 represent the noisy DNS data at noise levels $1/{\rm SNR} =$ 0.01, 0.1 and 1, respectively. (b) Results from the PCDRL model.}
\label{fig:F11}
\end{figure}

\begin{figure}
  \centerline{\includegraphics[scale=0.5]{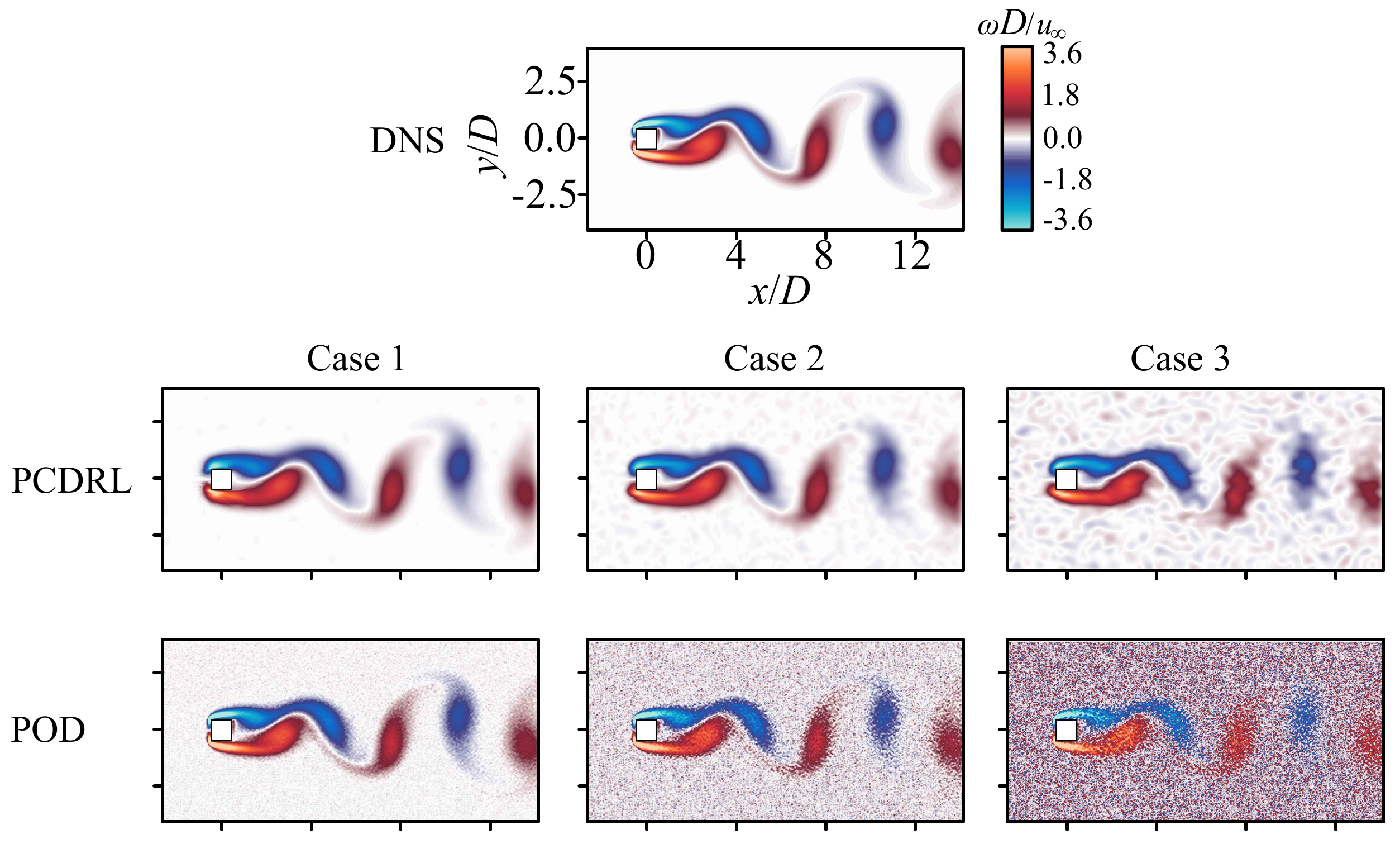}}
  \caption{ Reconstructed instantaneous vorticity field obtained from the DNS data using first ten POD modes. Cases 1, 2 and 3 represent the results of using noisy DNS data at noise levels $1/{\rm SNR} =$ 0.01, 0.1 and 1, respectively.}
\label{fig:F12}
\end{figure}

In this section, the accuracy of the results from the PCDRL model is examined in terms of flow decomposition. Firstly, the results of applying POD to the denoised data are compared with the POD results of the ground truth data. Figure~\ref{fig:F10} shows the contour plots of the leading POD modes for the vorticity field obtained from the DNS data. As can be observed from the figure, for the case of the highest noise level, i.e  1/SNR = 1, all the seven true leading modes can be recovered using the denoised data, while only three modes can be recovered using the noisy data and no distinguishable features can be seen for the other modes. Furthermore, the energy plots in Figure~\ref{fig:F11} represented by the normalised POD eigenvalues show that even for the case of the flow with the highest noise level, the energy contribution values of the POD modes are consistent with those obtained from the ground truth DNS data. As expected, the results from the noisy data reveal a different behaviour, especially for the cases of noise levels 1/SNR = 0.1 and 1. Figure~\ref{fig:F12} shows a reconstructed instantaneous vorticity field of the DNS data using the first ten POD modes. As shown in the figure, the result from PCDRL model reveals a commendable reconstruction accuracy as compared with the ground truth DNS results, whereas the obtained result from the noisy data indicates the limitation of POD to recover the flow with the right physics.

\begin{figure}
  \centerline{\includegraphics[scale=0.52]{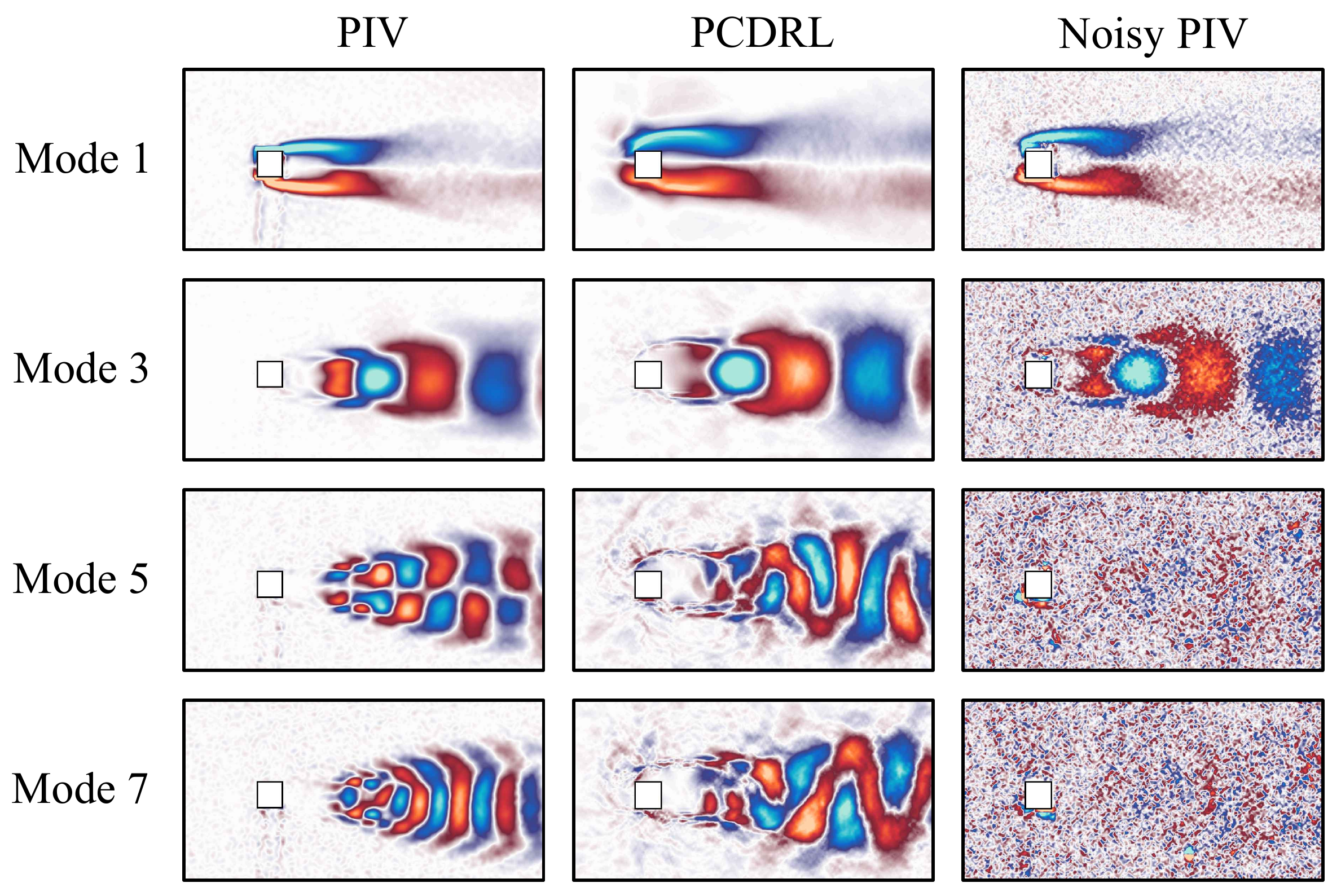}}
  \caption{Leading POD modes obtained from the PIV data. Results from clear PIV (left), PCDRL (middle), and noisy PIV data (right).}
\label{fig:F13}
\end{figure}

\begin{figure}
  \centerline{\includegraphics[scale=0.45]{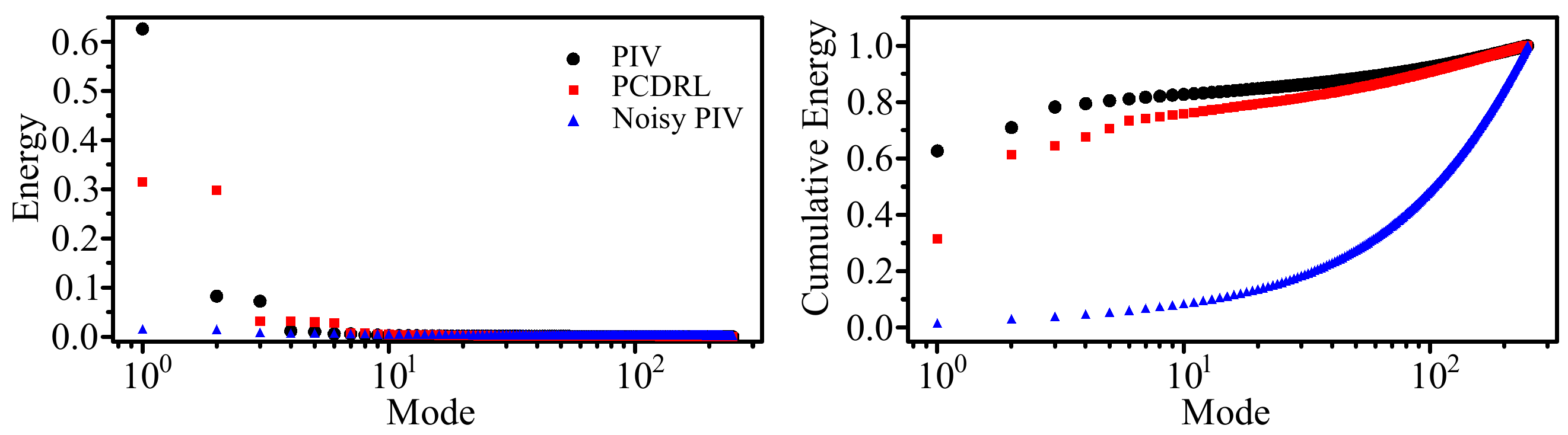}}
  \caption{Normlised energy (left) and cumulative energy (right) of the POD modes obtained from the PIV data.}
\label{fig:F14}
\end{figure}

\begin{figure}
  \centerline{\includegraphics[scale=0.5]{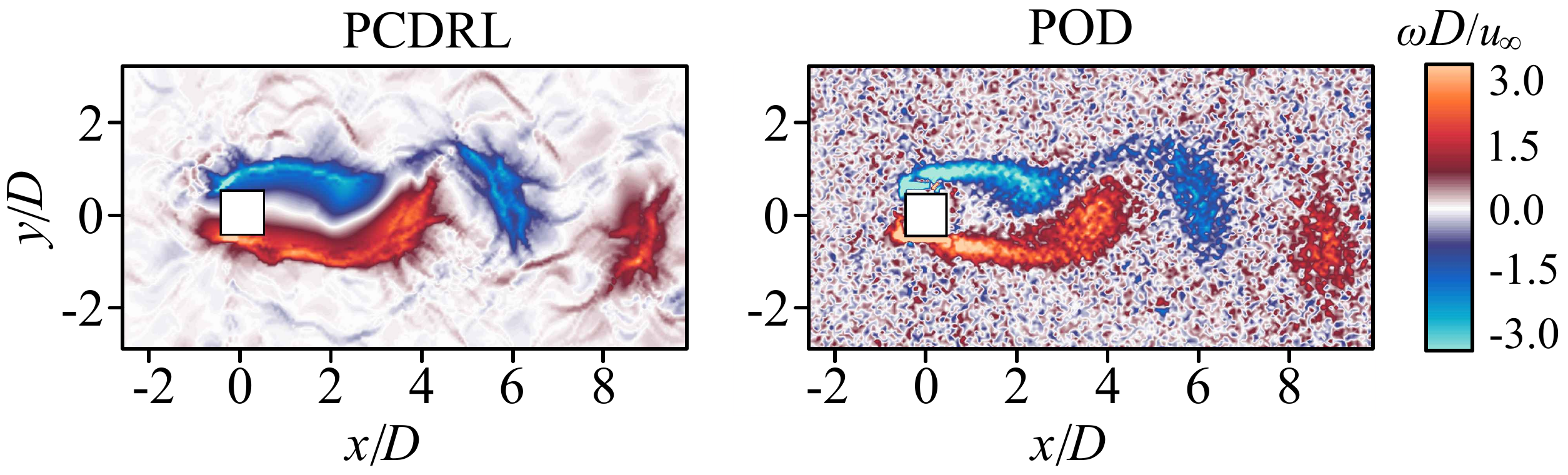}}
  \caption{Reconstructed instantaneous vorticity field obtained from the results of the PCDRL model (left) and noisy PIV data (right) using first ten POD modes.}
\label{fig:F15}
\end{figure}

Similar results can be obtained by applying the POD to the PIV data. As can be observed from Figure~\ref{fig:F13}, the seven leading POD modes obtained from the denoised PIV data are relatively consistent with the modes obtained from the clear PIV data, considering that the clean PIV data are obtained using a different experimental setup, whereas the noisy PIV data fails to recover the modes after the third mode. Notably, the shadow region is clearly shown in some of the modes obtained from the clear PIV data, whereas no such region can be seen in the modes obtained from the denoised data. This is consistent with results from Figure~\ref{fig:F9}(a). The results from Figure~\ref {fig:F14} further indicate the ability of the model to reconstruct the flow data with POD modes that generally have a behaviour similar to that of the clear PIV data. Furthermore, as shown in Figure~\ref{fig:F15}, the reconstructed instantaneous vorticity field using the first ten modes of the denoised data shows a realistic flow behaviour that is expected from the case of flow around a cylinder.

\begin{figure}
  \centerline{\includegraphics[scale=0.5]{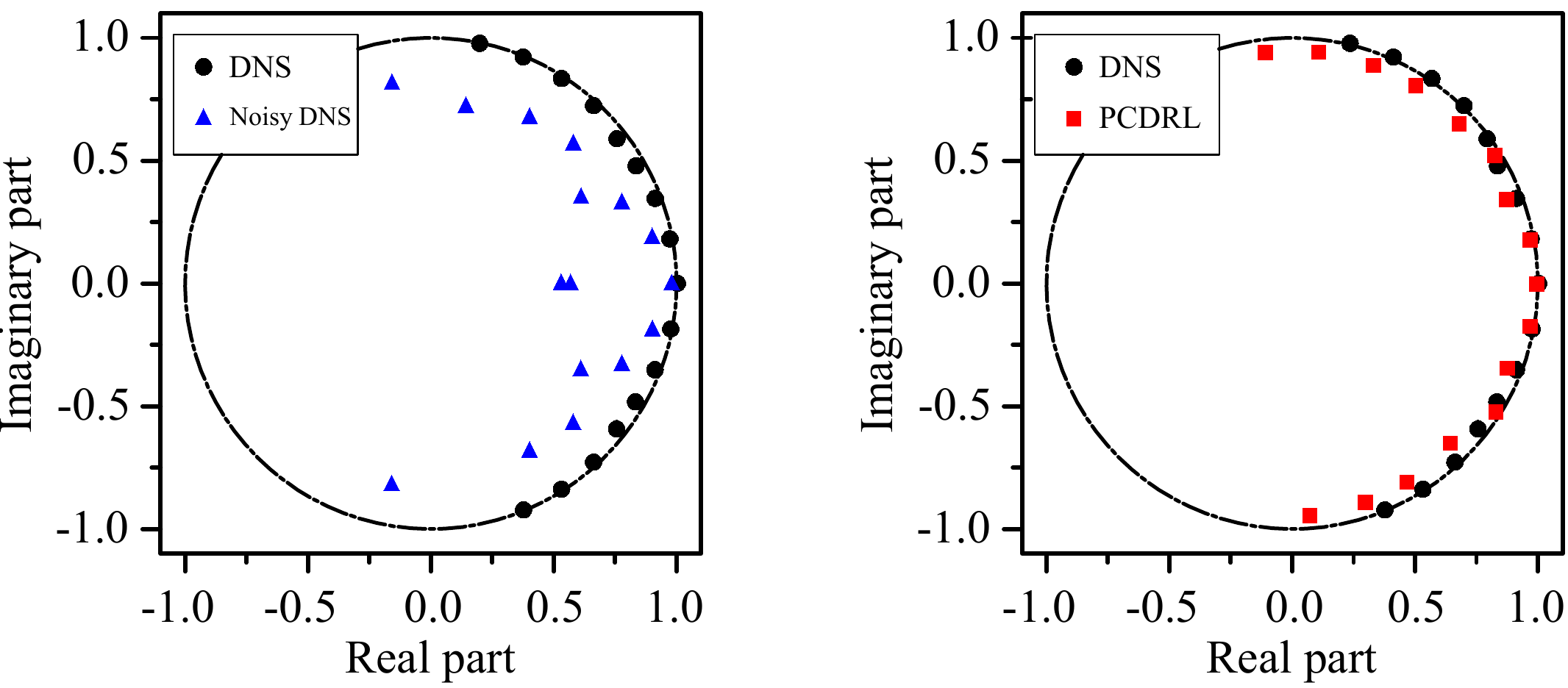}}
  \caption{DMD eigenvalues of the noisy DNS data at noise level 1/SNR = 1 (left) and the results from the PCDRL model (right) visualised on the unit circle.}
\label{fig:F16}
\end{figure}

To further investigate the dynamics of the denoised flow data, DMD is then applied to the flow data. As shown in Figure~\ref{fig:F16}, even in the case of the DNS data corrupted with the level of noise 1/SNR = 1, the DMD eigenvalues of the vorticity field show a behaviour close to that of the ground truth DNS data, whereas for the noisy data, the eigenvalues scatter inside the unit circle plot indicating a non-realistic behaviour of the system. 

\begin{figure}
  \centerline{\includegraphics[scale=0.5]{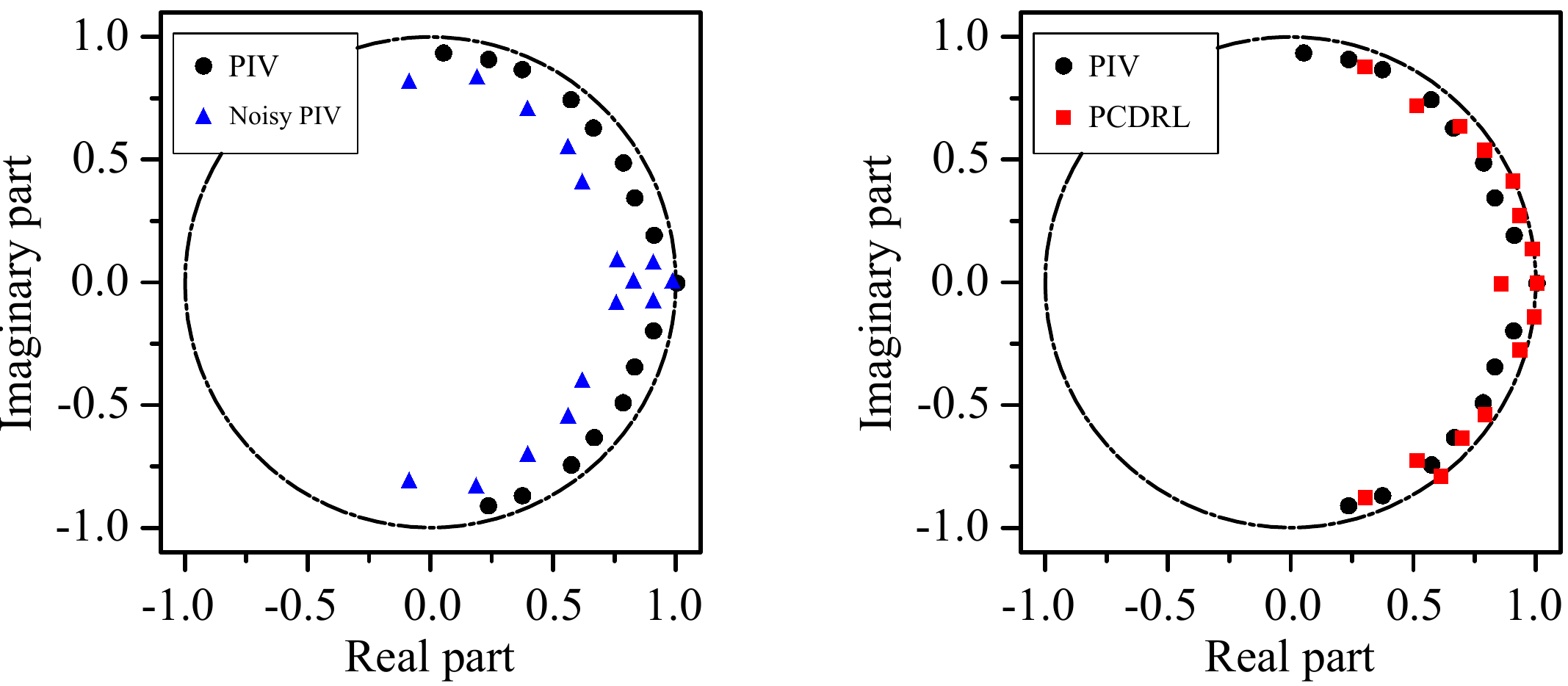}}
  \caption{DMD eigenvalues of the noisy PIV data (left) and the results from the PCDRL model (right) visualised on the unit circle.}
\label{fig:F17}
\end{figure}

As for the denoised PIV data, Figure~\ref{fig:F17} reveals that the eigenvalues also show a well-agreement with those from the clear PIV data. Notably, the leading DMD eigenvalues of the clear PIV data are not exactly located on the circumference of the unit circle as in the case of DNS data. This behaviour can be attributed to the fact that DMD is known to be sensitive to noise \cite{Shervin2014, Dawsonetal2016, Hematietal2017, Scherletal2020}, and unlike the DNS data, the clear PIV contains a relatively small level of noise, which can affect the flow decomposition.

\section{Conclusions} \label{Conclusions}
This study proposed a DRL-based method to reconstruct flow fields from noisy data. The PixelRL method is utilised to build the proposed PCDRL model, wherein an agent that applies actions represented by basic filters according to a local policy is assigned for each point in the flow. Hence, the proposed model is a multi-agent model. The physical constraints represented by the momentum equation and the pressure Poisson equation and the boundary conditions are utilised to build the reward function. Hence, the PCDRL model is label-training data-free, that is, target data are not required for the model training. Furthermore, the visualisation and interpretation of the model performance can be easily achieved due to the model setup.

The model performance was firstly investigated through DNS-based noisy data with three different noise levels. The instantaneous results and the flow statistics revealed a commendable reconstruction accuracy of the model. Furthermore, the spectral content of the flow was favourably recovered by the model with reduced accuracy with the increase in the noise level. Additionally, the reconstruction error showed relatively low values, indicating the general reconstruction accuracy of the model.

Real noisy and clear PIV data are utilised to examine the model performance. Herein, the model demonstrates its capability to recover the flow fields with the appropriate behaviour.

Furthermore, the accuracy of the denoised flow data from both DNS and PIV was investigated in terms of flow decomposition by means of POD and DMD. Herein, most of the leading POD modes that describe the main features (coherent structures) of the flow were successfully recovered with commendable accuracy and outperformed the results of directly applying POD to the noisy data. Additionally, the DMD eigenvalues obtained from the denoised flow data revealed a behaviour that is similar to that of true DMD modes. These results further indicate the model's ability to recover the flow data with most of the flow physics.

This study demonstrates that the combination of DRL, the physics of the flow, which is represented by the governing equations, and prior knowledge of the flow boundary conditions, can be effectively utilised to recover high-fidelity flow fields from noisy data. This approach can be further extended to be utlised in the reconstruction of three-dimensional turbulent flow fields. Herein, more sophisticated DRL models with more complex spatial filters are needed. Applying such models to flow reconstruction problems can result in a considerable reduction in the experimental and computational costs. 

\section*{Acknowledgments}\label{acknowledge}
This work was supported by `Human Resources Program in Energy Technology' of the Korea Institute of Energy Technology Evaluation and Planning (KETEP), granted financial resource from the Ministry of Trade, Industry \& Energy, Republic of Korea (no. 20214000000140). In addition, this work was supported by the National Research Foundation of Korea (NRF) grant funded by the Korea government (MSIP) (no. 2019R1I1A3A01058576)

\appendix

\section{Open-source code}\label{appA}
 The open-source library Pytorch 1.4.0 \cite{Paszkeetal2019} is utilised for the implementation of the model. The source code of the proposed PGDRL model is available at  \href{https://fluids.pusan.ac.kr/fluids/65416/subview.do?enc=Zm5jdDF8QEB8JTJGYmJzJTJGZmx1aWRzJTJGMTY1MzQlMkYxMTUwMTY1JTJGYXJ0Y2xWaWV3LmRvJTNGYmJzT3BlbldyZFNlcSUzRCUyNmlzVmlld01pbmUlM0RmYWxzZSUyNnNyY2hDb2x1bW4lM0QlMjZwYWdlJTNEMSUyNnNyY2hXcmQlM0QlMjZyZ3NCZ25kZVN0ciUzRCUyNmJic0NsU2VxJTNEJTI2cGFzc3dvcmQlM0QlMjZyZ3NFbmRkZVN0ciUzRCUyNg%3D%3D}{click here}.
 
\section{Asynchronous Advantage Actor–Critic (A3C)}\label{appB}

\begin{figure}[htbp]
\renewcommand{\thefigure}{B1}
  \centerline{\includegraphics[scale=0.1]{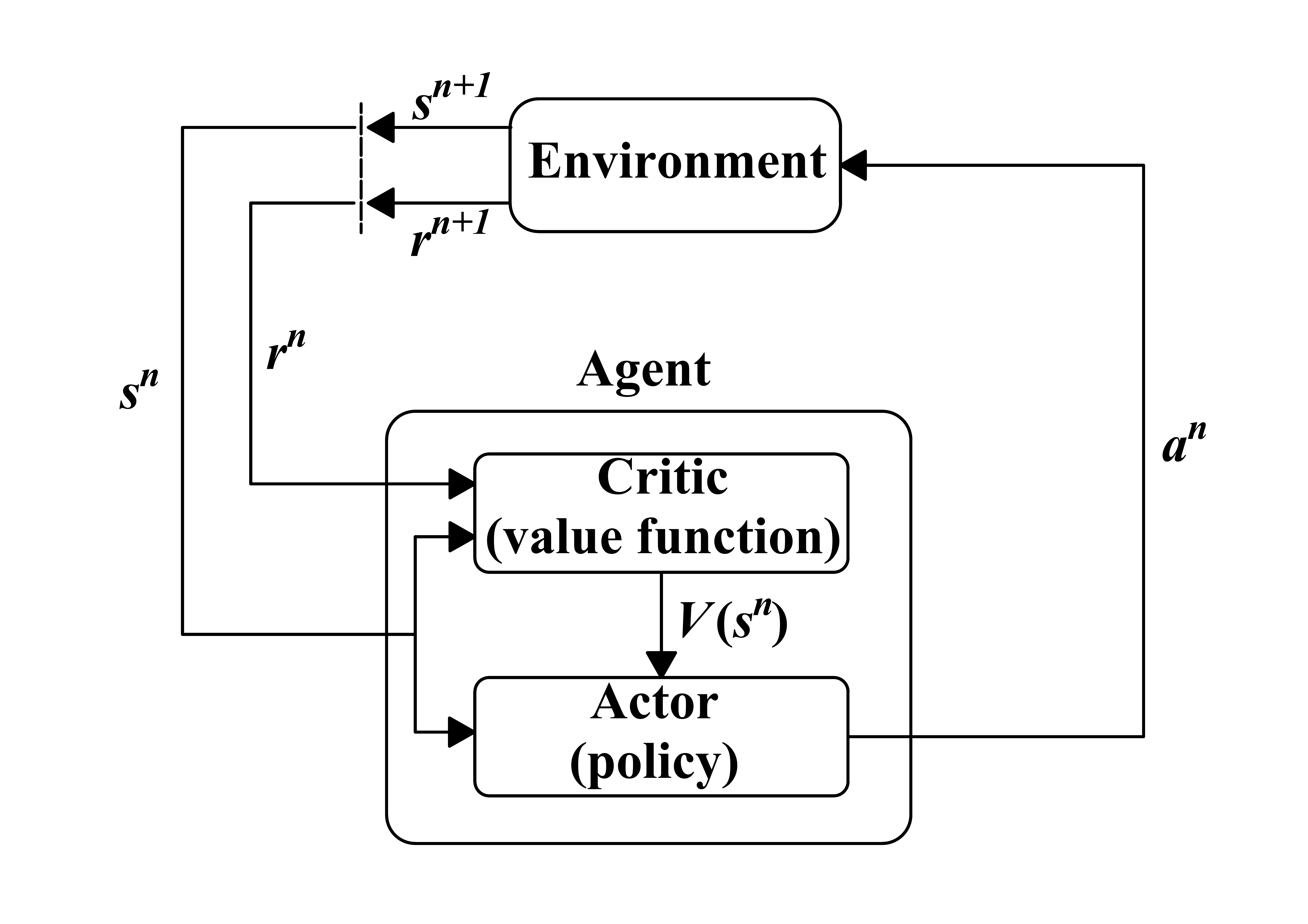}}
  \caption{Architecture of the actor–critic algorithm.}
 \label{fig:FB1}
\end{figure}

The asynchronous advantage actor–critic (A3C) \cite{Mnihetal2016} algorithm is applied in this paper. A3C is a variant of the actor–critic algorithm, which combines the policy- and value-based networks to improve its performance. Figure~\ref{fig:FB1} shows that the actor generates an action $a^n$ for the given state $s^n$ based on the current policy, whilst the critic provides the value function $V(s^n)$ to evaluate the effectiveness of the action.

Based on the understanding of the actor–critic algorithm, A3C also has two sub-networks: policy and value networks. Herein, $\theta_p$ and $\theta_v$ are used to represent the parameters of each network. The gradients of $\theta_p$ and $\theta_v$ can be calculated as follows:
\begin{equation} \tag{B1}\label{eqn:eqB1}
R^n = r^n + \gamma r^{n+1} + \cdots + \gamma^{(N-1)}r^N + \gamma^{(N)}V(s^N),
\end{equation}
\begin{equation} \tag{B2}\label{eqn:eqB2}
d\theta_v = \bigtriangledown_{\theta_v} (R^n - V(s^n))^2,
\end{equation}
\begin{equation} \tag{B3}\label{eqn:eqB3}
A(a^n,s^n) = R^n - V(s^n),
\end{equation}
\begin{equation} \tag{B4}\label{eqn:eqB4}
d\theta_p = -\bigtriangledown_{\theta_p} \log \pi (a^n,s^n) A(a^n,s^n),
\end{equation}

where $A(a^n,s^n)$ is the advantage.

\section{PixelRL}\label{appC}

\begin{figure}
\renewcommand{\thefigure}{C1}
  \centerline{\includegraphics[scale=0.14]{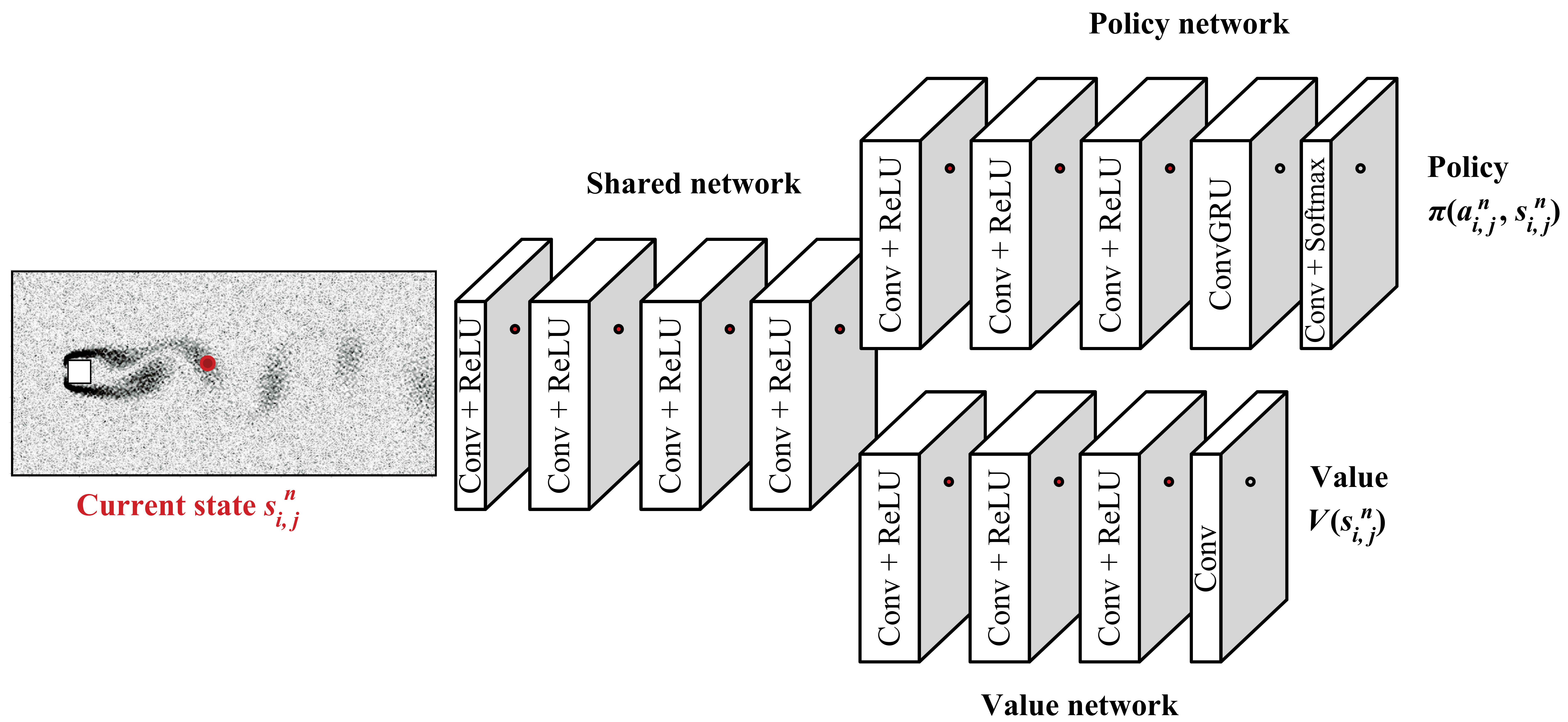}}
  \caption{Architecture of the fully convolutional A3C.}
\label{fig:FC1}
\end{figure}

A3C is modified in this study to a fully convolutional form \cite{Furutaetal2020}, and its architecture can be found in Figure~\ref{fig:FC1}. Through this approach, all the agents share the same parameters, which saves computational cost and trains the model more efficiently compared to the case where agents need to train their models individually. The size of the receptive field can also affect the performance of the CNN network, and a large receptive field can result in superior capture connections between points. Therefore, a receptive field ($3\times3$) is used in the architecture; that is, the outputs of the policy and value networks at a specific pixel will be affected by the pixel and its surrounding neighbor pixels. Figure~\ref{fig:FC1} shows that the input flow field data firstly pass through four convolutional and Leaky rectified linear unit (ReLU) \cite{Goodfellowetal2016} layers and are then inputted to the policy and value networks, respectively. The policy network comprises three convolutional layers with a ReLU activation function, a ConvGRU layer and a convolutional layer with a SoftMax activation function \cite{Goodfellowetal2016}, and its output is the policy. The first three layers of the value network are the same as the first three layers of the policy network, and the value function is finally obtained through the convolutional layer with a linear function.

The gradient of the parameters $\theta_p$ and $\theta_v$ is then defined on the basis of the architecture of the fully convolutional A3C.
\begin{equation} \tag{C1} \label{eqn:eqC1}
\textbf{\textit{R}}^n = \textbf{\textit{r}}^n + \gamma \textbf{\textit{W}} \ast \textbf{\textit{r}}^{n+1} + \cdots + \gamma^{(N-1)} \textbf{\textit{W}}^{(N-1)} \ast \textbf{\textit{r}}^N + \gamma^{(N)} \textbf{\textit{W}}^{(N)} \ast \textbf{\textit{V}}(\textbf{\textit{s}}^N),
\end{equation}
\begin{equation}  \tag{C2}\label{eqn:eqC2}
d\theta_v = \bigtriangledown_{\theta_v} \frac{1}{I\times J} \textbf{1}^\top \{ (\textbf{\textit{R}}^n - \textbf{\textit{V}}(\textbf{\textit{s}}^n)) \odot (\textbf{\textit{R}}^n - \textbf{\textit{V}}(\textbf{\textit{s}}^n))\} \textbf{1},
\end{equation}
\begin{equation}  \tag{C3}\label{eqn:eqC3}
\textbf{\textit{A}}(\textbf{\textit{a}}^n,\textbf{\textit{s}}^n) = \textbf{\textit{R}}^n - \textbf{\textit{V}}(\textbf{\textit{s}}^n),
\end{equation}
 \begin{equation}  \tag{C4}\label{eqn:eqC4}
d\theta_p = -\bigtriangledown_{\theta_p} \frac{1}{I\times J} \textbf{1}^\top \{ \log \bm{\pi} (\textbf{\textit{a}}^n,\textbf{\textit{s}}^n) \odot  \textbf{\textit{A}}(\textbf{\textit{a}}^n,\textbf{\textit{s}}^n) \} \textbf{1}.
\end{equation}
where $\textbf{\textit{R}}^n$, $\textbf{\textit{r}}^n$, $\textbf{\textit{V}}(\textbf{\textit{s}}^n)$, $\textbf{\textit{A}}(\textbf{\textit{a}}^n,\textbf{\textit{s}}^n)$ and $\bm{\pi} (\textbf{\textit{a}}^n,\textbf{\textit{s}}^n)$ are the matrices whose $(i, j)$-th elements are $R^n_{i,j}$, $r^n_{i,j}$, $V(s^n_{i,j})$, $A(a^n_{i,j},s^n_{i,j})$ and $\pi (a^n_{i,j},s^n_{i,j})$, respectively. $\ast$ is the convolution operator, $\textbf{1}$ is the all-ones vector and $\odot$ is the element-wise multiplication. $\textbf{\textit{W}}$ is a convolution filter weight, which is updated simultaneously with the $\theta_p$ and $\theta_v$ such that

 \begin{equation}  \tag{C5}\label{eqn:eqC5}
d\textbf{\textit{W}} = -\bigtriangledown_{\textbf{\textit{W}}} \frac{1}{I\times J} \textbf{1}^\top \{ \log \bm{\pi} (\textbf{\textit{a}}^n,\textbf{\textit{s}}^n) \odot  \textbf{\textit{A}}(\textbf{\textit{a}}^n,\textbf{\textit{s}}^n) \} \textbf{1} + \bigtriangledown_{\textbf{\textit{W}}} \frac{1}{I\times J} \textbf{1}^\top \{ (\textbf{\textit{R}}^n - \textbf{\textit{V}}(\textbf{\textit{s}}^n)) \odot (\textbf{\textit{R}}^n - \textbf{\textit{V}}(\textbf{\textit{s}}^n))\} \textbf{1}.
\end{equation}

Notably, after the agents complete their interaction with the environment, the gradients are acquired simultaneously, which means that the number of asynchronous threads is one; that is, A3C is equivalent to advantage actor–critic (A2C) in the current study \cite{Clementeetal2017}.

 \section{Denoising action set}\label{appD}

The action set for removing the noise of the flow fields is shown in Table~\ref{tab:TD1}. The agent can apply the following nine possible actions: do nothing, six classical image filters and plus/minus a $\textit{Scalar}$. The actions in this study are discrete and determined empirically. The table shows that the parameters $\sigma_c$, $\sigma_s$ and $\sigma$ represent the filter standard deviation in the color space, the coordinate space and the Gaussian kernel, respectively. The $\textit{Scalar}$ in the $8^{th}$ and $9^{th}$ actions is determined on the basis of the difference between the variance of the clear and noisy data.
\begin{table}
\renewcommand{\thetable}{D1}
 \caption{Action set for the denoising process}
  \centering
  \begin{tabular}{lll}
    \toprule
    Number     & Action type     & Parameters \\
    \midrule
    1     & Do nothing      & filter size = $5\times5$ \\
    2     & Box filter      & filter size = $5\times5$ \\
    3     & Bilateral filter1       & filter size = $5\times5$, $\sigma_c=1.0, \sigma_s=5.0$  \\
    4     & Bilateral filter2       & filter size = $5\times5$, $\sigma_c=0.1, \sigma_s=5.0$  \\
    5     & Median filter       & filter size = $5\times5$  \\
    6     & Gaussian filter1       & filter size = $5\times5$, $\sigma=1.5$  \\
    7     & Gaussian filter2       & filter size = $5\times5$, $\sigma=0.5$  \\
    8     & Value += $\textit{Scalar}$       &   \\
    9     & Value -= $\textit{Scalar}$       &    \\
    \bottomrule
  \end{tabular}
 \label{tab:TD1}
\end{table}\\

\bibliographystyle{unsrt}  
\bibliography{Arxiv_references-new}  

\end{document}